\documentclass[%
 aip,
 amsmath,amssymb,
preprint,%
 reprint,%
]{revtex4-1}

\usepackage{graphicx}
\usepackage[utf8]{inputenc}

\usepackage{rotating}
\usepackage{savesym}
\savesymbol{tablenum}  
\usepackage{siunitx}
\restoresymbol{SIX}{tablenum}

\usepackage{mathrsfs, amsmath, amssymb, amsthm, amsxtra, bm}
\usepackage{multirow}
\usepackage{hyperref}
\usepackage[nameinlink,capitalize,noabbrev]{cleveref}
\usepackage{todonotes}
\usepackage{svg}
\usepackage[section]{placeins}
\usepackage{pgfplotstable}
   \pgfplotsset{compat=1.18} 

\usepackage{colortbl}

\usepackage{markcmds}
\usepackage{seancmds}

\newcommand{\red}[1]{\textcolor{red}{\textsf{\textbf{#1}}}}
\newcommand{\blue}[1]{\textcolor{blue}{\textsf{\textbf{#1}}}}

\renewcommand{\red}[1]{#1}
\renewcommand{\blue}[1]{#1}

\def\printmathfont#1#2{%
  \ifcat\relax#2\relax
    \count0=#2%
  \else
    \count0=\mathcode`#2%
  \fi
  \divide\count0 256%
  \count1=\numexpr\count0/16\relax                
  \count0=\numexpr\count0 - 16*(\count0/16)\relax 
  \def\tempinfo##1{\xdef\temp{\fontname#1##1}}%
  $%
  \ifnum\count1=7\relax
    \ifnum\fam>0\relax
      \ifnum\fam<16\relax
        \tempinfo{\fam}%
      \else
        \tempinfo{\count0}%
      \fi
    \else
      \tempinfo{\count0}%
    \fi
  \else
    \tempinfo{\count0}%
  \fi
  $%
  \string#1 \string#2 ($#2$): \temp
}

\draft 

\begin{document}


\title{Direct power spectral density estimation from structure functions \red{without Fourier transforms}} 



\author{Mark A. Bishop}
\email{mark.bishop@vuw.ac.nz}
\affiliation{School of Chemical and Physical Sciences, 
Victoria University of Wellington,
Wellington 6012,
New Zealand}

\author{Sean Oughton}
\affiliation{Department of Mathematics,
University of Waikato,
Hamilton 3240,
New Zealand}

\author{Tulasi N. Parashar}
\affiliation{School of Chemical and Physical Sciences, 
Victoria University of Wellington,
Wellington 6012,
New Zealand}

\author{Yvette C. Perrott}
\affiliation{School of Chemical and Physical Sciences, 
Victoria University of Wellington,
Wellington 6012,
New Zealand}

\date{\today}

\begin{abstract}
Second-order structure functions and power spectral densities are popular tools in the study of statistical properties across scales, particularly for the analysis of turbulent flows. Although intimately related, analyses primarily use one or the other. We introduce a framework for estimating the power spectrum using the second-order structure function \red{without applying Fourier transforms} -- enabling one to take advantage of the real-space structure function calculations. We validate and showcase this method, comparing it to classical Fourier power spectrum estimates determined from analytical calculations, fractional Brownian motion, turbulence simulations, and space-physics and astrophysical observations of turbulence. We show that this method is able to robustly obtain the expected power law behaviour where we use turbulence ranges as test-cases.
\end{abstract}


\maketitle 

\section{Introduction}

In the analysis of complex physical systems -- particularly those exhibiting turbulent, fractal, or stochastic behaviour -- two key statistical tools are frequently employed: the power spectral density (PSD, which we will use interchangeably with power spectrum and spectrum) and the second order structure function (SF).

The power spectral density characterizes how the power (or variance) of a signal is distributed over some notion of \emph{scale} -- wavenumber or frequency. Power spectra are defined and often estimated through Fourier transform methods \citep{Bracewell86}. These methods come in many variations, \eg{}the periodogram \citep{Schuster98}, Blackman-Tukey \citep{Blackman.etal60}, Welch's method, and the Lomb-Scargle periodogram \citep{VanderPlas18}, \etc{}. These classical/non-parametric methods have been developed over decades and have their respective advantages, disadvantages, and use-cases depending on the behaviour of the data that are being examined \citep{Stoica.Moses05, Babu.Stoica10, Thomson.Haley14}. Some of these use-cases refer to gapped or non-uniformly sampled data which breaks the classic Nyquist aliasing relation and leads to degraded performance on spectral estimation \cite{Maciejewski.etal09}. Vast amounts of research has been performed in comparing and quantifying the influences of gapped and non-uniformly sampled data for different methods of PSD estimation through \eg{}interpolation of the data in gaps, or window/tapering the data \cite{Hocke.Kampfer09, Smith-Boughner.Constable12, Munteanu.etal16, Jahanjooy.etal16, Roelens.etal17, Chave19}.

Alternatively, the structure function -- specifically, the second-order structure function -- measures the average squared difference between the signal at two points separated by a given lag. This lag separation is also some notion of \emph{scale} -- in the time or spatial domain. The structure function is often estimated directly from the observed data where lag separations are available \ie{}values are based off pairs in the data, so if one or both values in a pair are missing, it can simply be ignored \cite{Burlaga91, Oliver.Webster14, Oliver.Webster15, Sadhukhan.etal21, Gatuzz.etal23}.

Though PSDs and SFs operate in different domains (frequency/wavenumber versus time/space), under assumptions of stationarity the SF is mathematically related to the PSD by an integration of the PSD \cite{Monin.Iaglom75, Frisch95, Pope00, Emmanoulopoulos.etal10}. In the case of a stationary, homogeneous process with a PSD that has a range of wavenumbers that is a power-law $\mathcal{E}_{D}(k) \propto k^{-\beta}$, the second-order SF $\overline{S}_{D}(\ell) \propto \ell^{\eta}$ is expected to scale with $\eta = \beta - 1$ within specific bounds $1 < \beta < 3$.

In turbulence studies power-law scaling is a key signature of energy cascades for both the PSD \cite{Kolmogorov41} and the SF \cite{Kolmogorov91}. The spectral slopes in either domain have been used to estimate the turbulent kinetic energy dissipation rates across various environments including the ocean \citep{Lorke.Wuest05, Wiles.etal06}, the atmosphere \citep{Cohn95}, and chemical processes \citep{Wang.etal21a}. The power-law scaling has been used to understand the role turbulence plays in promoting and inhibiting star-formation in the interstellar medium with both the PSD and SF providing complementary statistics about energy cascades via spectral slopes, and transitional scales via changes in slopes \cite{Brunt.Heyer02, Boldyrev.etal02, Padoan.etal03, Brunt.etal03, Federrath.etal21}.

Since the earliest of space missions, the PSD and SF spectral estimates have been used to measure the properties of turbulence in the solar wind \cite{Coleman68, Matthaeus.Goldstein82}, a ``laboratory'' for plasma turbulence experiments \cite{Marsch91, Tu.Marsch94, Bruno.Carbone13, Oughton.etal15, Chen16, Verscharen.etal19, Smith.Vasquez21, Fraternale.etal22}. This is an astrophysical context that often provides \emph{in-situ} measurements, typically as 1-dimensional time-series data. SFs are frequently employed to mitigate the effect of data gaps and telemetry dropouts with accurate spectral estimation claimed for missing fractions of up to 68\% \cite{Gallana.etal16, Burger.McKee23, Dorseth.etal24a}. Some work has shown that the SF is susceptible to missing data \cite{Emmanoulopoulos.etal10, Fraternale.etal19}, but it has been suggested that this can be mitigated using empirically derived correction factors \cite{Wrench.Parashar24}.

Beyond solar wind studies, two-dimensional telescope observations of line emissions of the interstellar and intracluster mediums are used to infer the velocity structure of astrophysical fluids \cite{Koch.etal19, Burkhart21}. However, these images often consist of irregularly shaped or non-uniformly distributed binned pixels, which makes traditional Fourier-based PSD estimation difficult. Therefore, SF analysis is often used \cite{Li.etal20, Ganguly.etal23, Gatuzz.etal23}.

Similarly, in two-dimensional X-ray observations of the intracluster medium, the surface brightness maps often contain gaps caused by observational constraints or the removal of contaminating sources such as galaxies \cite{Churazov.etal12, Zhuravleva.etal19, Romero.etal23}. These gaps introduce aliasing artifacts in Fourier PSD estimates, prompting the use of spatial-domain techniques \cite{Stutzki.etal98, Bensch.etal01, Ossenkopf.etal08, Arevalo.etal12}. Full spectral characterization can still be useful as correcting for instrumental or observational noise is often done in Fourier space \cite{Churazov.etal12} whereas uncertainties are better understood in the spatial domain \cite{Emmanoulopoulos.etal10, Clerc.etal19, Cucchetti.etal19}.



In this paper, we discuss a method for directly estimating the power spectrum using second-order structure functions. Practically speaking, this method uses lag-space computations, which can be advantageous when Fourier-space computations may introduce aliasing problems \cite{Jing05, Arevalo.etal12, Delsuc.OConnor24}.

Previous studies have used approximate variants of this approach, including in the analysis of solar wind time-series data \cite{Chasapis.etal17, Parashar.etal18, Chhiber.etal18, Thepthong.etal23} and laboratory turbulence experiments \cite{Huang.etal09, Huang.etal10}. Our work extends these efforts by using a more rigorous derivation \cite{Squire.Hopkins17, Thepthong.etal23}, including corrections for systematic biases that were previously unrecognized. Additionally, we provide a detailed investigation of the mathematical relationship between the lag distance and Fourier wavenumber and provide extensions of common values to arbitrary dimension.

The remainder of this paper starts with detailing the calculations required to obtain the SF to PSD relation (\autoref{sec:method}). Next, \autoref{sec:amplitude_and_wavenumber_biases} discusses the biases and factors associated with approximating the PSD with the structure function. This is accomplished through analytical work and examples with additional analytical work in Appendix \ref{app:additional_analytics}. \autoref{sec:implementation_testing} introduces details regarding the practical implementation of, and validates the performance on fractional Brownian motion (fBm) fields with comparison to the analytical scenarios of the previous section. Finally, \autoref{sec:examples} showcases the method with high-resolution real-world datasets/observations.

\section{Method}
\label{sec:method}

This section outlines the approach we take to obtain an
        \emph{equivalent spectrum},
by which we mean a quantity calculated using structure functions (or
similar quantities)
that---without use of a Fourier transform---can be interpreted as an
approximation to the actual spectrum.
Similar approaches and formulae have been presented previously
\cite{Davidson.Pearson05, Huang.etal09, Hamba15, Squire.Hopkins17,
  Thepthong.etal23}.

We begin by considering a zero-mean scalar field,
        $ s_D ( \vec{x} ) $,
for which measurements are available in a Euclidean space of dimension
        $ D $,
with $\vec{x} \in \mathbb{R}^D$
the vector of spatial coordinates.
\blue{An important distinction here is
that $D$ is not necessarily the dimension of the
system itself ($D_s$, say).
Indeed, the latter will often be larger than $D$ since measurement
techniques are frequently restricted to 1D or 2D samples of 3D systems.
We assume that the statistics (e.g., moments) of
        $ s_D(\vct(x))$
are homogeneous.}


Recall that the angle-averaged second-order structure function,
\begin{align}
    \label{eqn:averaged_SF_defn}
    \barS = \avg{ \abs{s_D(\vec{x} + \vec{\ell}) - s_D(\vec{x})}^2 } ,
\end{align}
is related to the angle-averaged autocorrelation function,
\begin{align}
    \label{eqn:averaged_acf_defn}
    \barR = \avg{ s_D(\vec{x}) s_D(\vec{x} + \vec{\ell}) } ,
\end{align}
by the standard result
\begin{align}
    \label{eqn:sf_variance_acf}
    \barS = 2 \overline{R}_{D}(0) - 2 \overline{R}_D(\ell),
\end{align}
where the overbar and $\avg{\cdot}$ indicate averaging over two
things: position $\vec{x}$,
and the direction of the lag vector $\vec{\ell}$ for fixed magnitude $\ell$. Homogeneity means that quantities so averaged do
not depend on the (absolute) positions $ \vct{x} $.


Letting $ \ell \rightarrow \infty$, we have
\begin{align}
    \label{eqn:sf_variance}
    \overline{S}_{D}(\infty) &= 2 \overline{R}_D(0) = 2 \avg{s_{D}(\vec{x})^2},
\end{align}
since
        $ \lim_{\ell \rightarrow \infty} \overline{R}_D(\ell) = 0 $.
As this is essentially the `energy' associated with  $ s_D $,
we may make
use of Parseval's theorem to relate it to the (Fourier space) integral
of an angle-integrated energy spectrum, $\Earg$,
\begin{align}
    \label{eqn:sf_fundamental_theorem}
    \overline{S}_{D}(\infty)
        = 2 \int_0^\infty \Earg \, \d k
        = \int_0^\infty \frac{\d \barS}{\d \ell} \, \d \ell,
\end{align}
where the rightmost form follows from the fundamental theorem of
calculus and
        $ \overline{S}_{D}(0) = 0 $.
\blue{The definition
   of $\Earg$---the angle-integrated spectrum
   available when the measurement space dimension is $D$---is important
   and will be discussed shortly.}

Introducing the equivalent wavenumber\footnote{In our convention,
  there is a factor of $2\pi\,\mathrm{radian}$ in the wavenumber:
  $k$ and $\ke$ are really the \emph{angular} wavenumbers.
  Since $b$ is a dimensionless constant,
  when converting $b/\ell$ to $\ke$ there is also an
  implicit multiplication of $1\,\mathrm{radian}$. As
  $\mathrm{radians}$ are really dimension-less, there is no real
  consequence other than resolving confusing perspectives regarding
  angular units.}
\begin{align}
    \label{eqn:equivalent_wavenumber}
    \ke = \frac{b}{\ell},
\end{align}
where $b$ is a conversion factor to be defined later, and using it to make a
change of variable in \autoref{eqn:sf_fundamental_theorem}
yields
\begin{align}
    \label{eqn:esf_integrals}
      \int_0^\infty \frac{\ell^2}{b} \frac{\d \barS}{\d \ell} \, \d \ke
        =
    2 \int_0^\infty \Earg \, \d k.
\end{align}

\red{\autoref{eqn:esf_integrals} is exact, although of course it does
not mean that the integrands are equal.
Nonetheless, we now assume that there may be some scale range(s)
over which the integrands are approximately equal,
and define the \emph{equivalent spectrum} by,}
\begin{align}
    \label{eqn:equivalent_spectrum}  
    \ES \bracket{k_{\mathrm{e}}}
      & =
        \left.
        \frac{1}{2} \frac{1}{b} \ell^2
        \frac{\d \overline{S}_{D}(\ell)} {\d  \ell}
        \right|_{\ell = b/k_e}
  .
\end{align}
When the near equality of the integrands is good, we will then have
\begin{align}
    \label{eqn:equivalent_spectrum_relationship}
    \ES( \ke ) \approx
        \mathcal{B} \mathcal{E}_{D} (\ke ) ,
\end{align}
\red{where the bias $\mathcal{B}$ is discussed below.
 The relationship
 to \emph{signature functions} \cite{Davidson15} is discussed
 in \autoref{app:signature_function_relations}.
}

\blue{To reiterate: for homogeneous turbulence
  \autoref{eqn:esf_integrals} is
  exact---however, it of course does \emph{not} mean that the integrands
  (used in inferring \autoref{eqn:equivalent_spectrum}) are pointwise
  equal. For example, the following definite integrals are equal:
  $\int_0^{1} x \, \d x = \frac{1}{2} \int_{0}^{1} \d x$ and yet their
  integrands are not. Nonetheless we will demonstrate there are
  circumstances under which there is an equivalence/proportionality of
  the integrands, at least for some range of  $ \ke $.
  These circumstances are explored analytically in
  \autoref{sec:amplitude_and_wavenumber_biases} and empirically
  validated in \autoref{sec:implementation_testing}, making use of
  several model spectra that are relevant to turbulence problems.
  As we shall see, that analysis indicates that the methodology can be
  useful in various applications. However, care is needed.
  As $b$ and $ \mathcal{B} $
  typically vary with the application, we urge readers
  to individually examine the
  systematic biases for their specific use-cases.}

As a technical matter, $\ESarg$ is a \emph{biased} estimator that approximates
the actual spectrum $\Earg$
by allowing for a possible deviation in amplitude by the factor
        $ \mathcal{B}$ and
a possible shift of the wavenumber argument.
Determining suitable values for
        $ \mathcal{B}$ and $b$
is a key step in obtaining a useful and accurate equivalent spectrum and
these factors are discussed in detail in the next section.

\blue{Let us return to the definition of $\Earg $.
  We will need to distinguish between several types of spectra: modal,
  angle-averaged, and reduced. See Appendix \ref{app:d_spectra} for
  definitions.
  (Note that when $D=3$ some works call the angle-averaged spectra the
  3D spectrum \cite{Tennekes.Lumley72, Davidson15}. Herein, we will
  use the less ambiguous terms angle-averaged or omni-directional.)
  Suppose that we have access to realizations of the
  full continuum of values of $s_D(\vct{x})$ in the $D$-dimensional
  space in which measurements occur
     (e.g., $D=2$ because of line-of-sight integration,
     or $D=1$ for observations made by a spacecraft).
  Using these realizations, one can construct the correlation function
  for all lags $\vct{\ell}$ and Fourier transform it to obtain a modal
  spectrum, $ E_D (\vct{k}_D) $; see \autoref{eqn:fourier_spectrum_defn}. Employing appropriate polar coordinates,
  this can be integrated over the $D-1$ angles to get the $\Earg $
  that appears in
    \autoref{eqn:sf_fundamental_theorem}--\autoref{eqn:equivalent_spectrum_relationship}.
    It is this quantity that the equivalent spectrum is an
    approximation to.
}

\blue{Frequently, the measurements will be made in 1D or 2D,
   while the system itself will be 3D.  For those situations, the modal and
   angle-integrated spectra for the \emph{system}
   will differ from the spectra
        $ E_D (\vct{k}_D) $ and $\Earg $
   obtainable from the measurement dataset.
   Examples are given in
        Appendix
        \ref{sec:sliced_and_projected_data}. Moreover, as is well known,
   except in special cases (e.g., high symmetry),
   it is not usually possible to obtain the
   angle-integrated
   spectrum associated with the full system, $\mathcal{E}^\text{sys} (k) $, say,
   from  $\Earg $.  }

Note also that
the definitions and formulae provided here are also valid when
 $ s_D (\vct{x}) $ is replaced by a vector field
 $ \vec{v}_{D}(\vec{x}) $,
in which case the absolute value becomes the
vector norm and the scalar multiplication in \autoref{eqn:averaged_acf_defn}
becomes the Euclidean dot product.

\section{Amplitude $\mathcal{B}$ \& Wavenumber $b$ Bias}
\label{sec:amplitude_and_wavenumber_biases}
\autoref{eqn:equivalent_wavenumber} and \autoref{eqn:equivalent_spectrum} have introduced somewhat arbitrary factors ($\mathcal{B}$ and $b$) that can be thought of as corresponding to an amplitude bias ($\mathcal{B}$) and wavenumber bias ($b$). Here, bias is used to mean a systematic correction is needed. The amplitude bias is associated with correcting the ``power'' of $\ESarg$ and will often be considered as the ratio of the genuine Fourier spectrum $\Earg$ to the structure function based estimate $\ESarg$. The wavenumber bias is associated with converting the lag-scale $\ell$ to the Fourier wavenumber $k$.

This section describes the process for investigating the appropriate amplitude factor $\mathcal{B}$ and wavenumber factor $b$ which could be used in future work, and developed further for specific use cases. We propose formulae and approximations for $\mathcal{B}$ and $b$ that result in adequate estimations for most turbulence spectra. These formulae should be applicable for many other scale-dependent phenomena assuming their spectra are similar to the ones we model.

For reference, \autoref{tab:symbols} provides a list of symbols and their definitions used throughout this paper.

\begin{table*}
    \centering
        \begin{tabular}{ c l l }
            \hline
            \hline
            Name & Symbol & Definition(s) \\
            \hline
            \hline
            \rule{0pt}{4ex}
            Angle-averaged second-order structure function & $\barS$ & \autoref{eqn:averaged_SF_defn}, \autoref{eqn:spectral_representation}\\
            Angle-averaged autocorrelation function & $\barR$ & \autoref{eqn:averaged_acf_defn}\\
            Angle-integrated spectrum & $\Earg$ & \autoref{eqn:sf_fundamental_theorem}, \autoref{eqn:integrated_spectrum_from_acf}\\
            \hline
            Equivalent wavenumber & $\ke$ & \autoref{eqn:equivalent_wavenumber}\\
            Amplitude bias correction factor & $\mathcal{B}$ & \autoref{sec:amplitude_and_wavenumber_biases}\\
            Wavenumber bias correction factor & $b$ & \autoref{sec:amplitude_and_wavenumber_biases}\\
            Equivalent spectrum & $\ESarg$ & \autoref{eqn:equivalent_spectrum}, \autoref{eqn:equivalent_spectrum_via_acf}, \autoref{eqn:equiv_spectrum_filter}\\
            Equivalent spectrum (power law approx.) & $\widetilde{\mathcal{P}}_{D}^{S}(\ke)$ & \autoref{eqn:powerlaw_ESF_approx}\\
            Local power law slope for $\ESarg$ & $\Delta \tilde{\beta}(\ke)$ & \autoref{eqn:local_powerlaw_est}\\
            \hline
            Effective energy-containing scale for $\Earg$ & $\kp$ & \autoref{eqn:kp}\\
            Effective energy-containing scale for $\ESarg$ & $\kep$ & \autoref{eqn:kep}\\
            \hline
            Estimated $b$ factor at $\kp$ & $\bpest$ & \autoref{eqn:bp_est}\\
            Empirical $b$ factor at $\kp$ & $\bpemp$ & \autoref{eqn:empirical_b_estimate}\\
            \hline
            Analytical bias for a pure power law function & $\Bpow$ & \autoref{eqn:analytical_equivalent}\\
            Analytical bias for the exp. model spectrum & $\Bexp$ & \autoref{eqn:meijerG_bias}\\
            Analytical bias for the exp. model using $\widetilde{\mathcal{P}}_{D}^{S}(\ke)$ & $\widetilde{B}_{D}^{\mathrm{exp-pow}}$ & \autoref{eqn:pow_approximation_exp_bias}\\
        \end{tabular}
    \caption{\textbf{Table of frequently used symbols in this paper}.}
    \label{tab:symbols}
\end{table*}

\subsection{Pure Power Law Spectrum}
\label{sec:pure_powerlaw_bias}
As has been extensively discussed in the literature, a relationship between the structure function and the amplitudes of the power spectrum can be calculated for pure power law spectra \cite{Pope00, Frisch95}. Our equivalent spectrum approach also has a bias $\mathcal{B}$ which is readily calculated for a pure power law. Using the following pure power law spectrum with $1 < \beta < 3$,
\begin{align}
    \label{eqn:powerlaw_spectrum}
    \Earg = A k^{-\beta},
\end{align}
we can calculate the structure function using its exact spectral representation \citep{Monin.Iaglom75, Emmanoulopoulos.etal10},
\begin{align}
    \label{eqn:spectral_representation}
   \barS &= 2 \int_0^\infty \sqbracket{1 - \mathcal{T}_D(k \ell)} \Earg \d k,
\end{align}
\red{where $\mathcal{T}_{D}(k\ell)$ is given by \autoref{eqn:transform}.}

The equivalent spectrum corresponding to the pure power-law $\mathcal{E}_{D}(k)$ is analytically evaluated using \autoref{eqn:spectral_representation} as
\begin{align}
    \label{eqn:analytical_equivalent}
    \ESarg &= \underbrace{\bracket{\frac{2}{b}}^{1-\beta} \frac{\Gamma\bracket{\frac{D}{2}} \Gamma\bracket{\frac{3-\beta}{2}}}{\Gamma\bracket{\frac{\beta+D-1}{2}}}}_{\Bpow} \underbrace{\vphantom{\frac{\Gamma\bracket{\frac{D}{2}} \Gamma\bracket{\frac{3-\beta}{2}}}{\Gamma\bracket{\frac{\beta+D-1}{2}}}} \sqbracket{A k_{\mathrm{e}}^{-\beta}}}_{\E(\ke)}.
\end{align}
As indicated by the square brackets, clearly $\ESarg$ contains $\E(k=k_{\mathrm{e}})$. However, there exist additional coefficients that depend on $b$, $D$, and $\beta$. These coefficients (denoted $\Bpow$ in \autoref{eqn:analytical_equivalent}) form a natural bias in the spectral estimation \ie{}$\ESarg$ has a different amplitude than $\E(k=k_{\mathrm{e}})$. \red{Such amplitude biases have previously been investigated in the \emph{signature function} literature \cite{Davidson.Pearson05, Davidson15} (see, Appendix \ref{app:signature_function_relations} for more details).}

\begin{figure}
    \centering
    \includegraphics[scale=1]{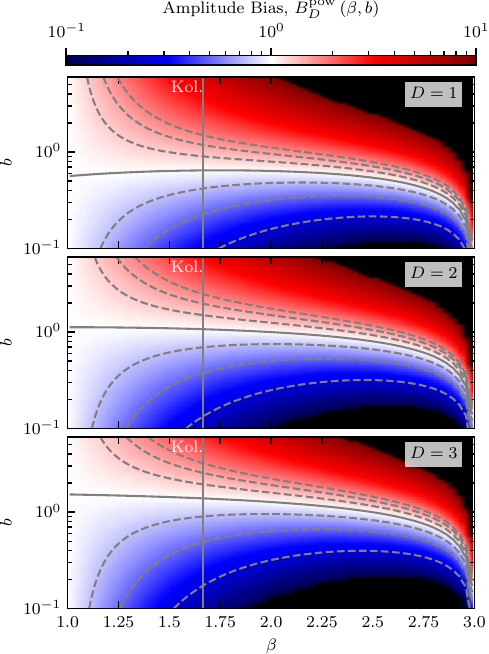}
    \caption{The amplitude bias \autoref{eqn:powerlaw_bias} for a pure power law angle-integrated spectrum \autoref{eqn:powerlaw_spectrum}, for Euclidean dimensions $D=1$, $D=2$, and $D=3$. The line contours correspond to linear spacings of 0.25 starting from $\Bpow=0.25$ to $1.75$ with $\Bpow=1$ represented by the solid gray line. The solid vertical gray line represents the (Kolmogorov) power law slope of $5/3$.}%
    \label{fig:numerical_B_powerlaw_estimate}%
\end{figure}

\begin{figure}
    \centering
    \includegraphics[scale=1]{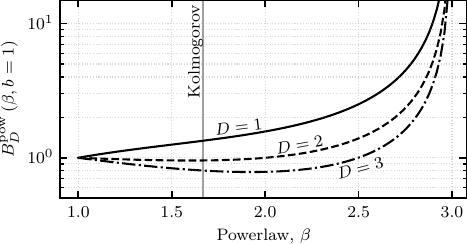}
    \caption{The amplitude bias \autoref{eqn:powerlaw_bias} with fixed $b=1$ for a pure power law angle-integrated spectrum \autoref{eqn:powerlaw_spectrum}, for Euclidean dimensions $D=1$ (dashed), $D=2$ (dotted), and $D=3$ (dash-dotted). The solid vertical gray line represents the (Kolmogorov) power law slope of $-5/3$.}%
    \label{fig:analytical_amplitude_bias}%
\end{figure}

We now emphasize two scenarios that can be deduced from the above equations:
\begin{itemize}
    \item \textbf{$b$ is a function of the spectral index $\beta$.}\\
    We could determine a $b$ to equate \autoref{eqn:analytical_equivalent} to \autoref{eqn:powerlaw_spectrum}, \ie{}$\Bpowarg = 1$ implies
    \begin{align} 
        \label{eqn:analytical_b_for_powerlaw}
        b = b^{\mathrm{pow}}_{D}(\beta) \equiv 2 \sqbracket{ \frac{\Gamma\bracket{\frac{D}{2}} \Gamma\bracket{\frac{3-\beta}{2}}}{\Gamma\bracket{\frac{\beta+D-1}{2}}} }^{1/(1 - \beta)}.
    \end{align}
    We can clearly see that the value of $b$ needed to equate \autoref{eqn:analytical_equivalent} to \autoref{eqn:powerlaw_spectrum} changes with $\beta$ and $D$ (see the $\Bpowarg = 1$ contour in \autoref{fig:numerical_B_powerlaw_estimate}) -- the source of this result is discussed in \autoref{sec:filter_function}.

    If we were to choose a $b$ dependent on the power law index $\beta$ we would complicate the relationship between a wavenumber $k$ and lag distance $\ell$. Not to mention, practically determining $b$ would require \emph{a priori} spectral slope knowledge.

    In this case, \emph{there is no universal $b$}.

    \item \textbf{$b$ is a constant.}\\
    It would be more natural to fix $b = \text{const.}$ to obtain a physically motivated relationship $\ke = b/\ell$ describing the connection between the Fourier wavenumber and a real-space (lag) scale.

    Equivalence of \autoref{eqn:analytical_equivalent} and \autoref{eqn:powerlaw_spectrum} requires that
    \begin{align}
        \label{eqn:powerlaw_bias}
        \mathcal{B} = \Bpowarg \equiv \bracket{\frac{2}{b}}^{1 - \beta} \frac{\Gamma\bracket{\frac{D}{2}} \Gamma\bracket{\frac{3-\beta}{2}}}{\Gamma\bracket{\frac{\beta+D-1}{2}}}.
    \end{align}
    In \autoref{fig:numerical_B_powerlaw_estimate}, we display a visualization of \autoref{eqn:powerlaw_bias} showing how the bias factor $\Bpowarg$ behaves as a function of $b$ and $\beta$ for $D=1,\,2,\,3$. In other words, $b$ does not influence the spectral power-law -- it is possible to obtain the correct power-law slope for any reasonable, constant choice of $b$. Any discrepancy between the chosen $b$ and \autoref{eqn:analytical_b_for_powerlaw} is simply absorbed into the systematic bias \autoref{eqn:powerlaw_bias}. However, with different values employed in the literature, unfortunately, a constant $b$ factor is a great source of ambiguity -- with values of $1,\,\pi,$ or $2\pi$ commonly used \cite{Davidson.Pearson05, Davidson15, Colbrook.etal17, Squire.Hopkins17, Federrath.etal21, Thepthong.etal23}. This ambiguity is highlighted in the next section.

    For now, if we assert $b=1$, then as \autoref{fig:analytical_amplitude_bias} shows, $\Bpow$ is low (for $D=2,3$) to modest (for $D=1$) for reasonable $\beta$ values (\eg{}Kolmogorov $\beta \approx 5/3$). This explains some of the success in previous solar wind ($D=1$) analyses \cite{Chhiber.etal18, Thepthong.etal23} which effectively applied $\ESarg$ with $b=1$ \red{(see Appendix \ref{sec:sliced_and_projected_data} for further details)}. Also notable, is the bias for $D=2$: it is close to unity for $1 < \beta \lesssim 2$ (\ie{}nearly no bias), suggesting that this is an effective method for estimating power spectral densities of observations of turbulence (telescope images) in astrophysical plasmas such as the intracluster medium (ICM) of galaxy clusters \citep{Simionescu.etal19}, and the interstellar medium (ISM) \citep{Burkhart21}. However, we note that telescope images are 2-dimensional projections of 3-dimensional physical structures, and therefore do not correspond to truly $D=2$ measurements \red{(see Appendix \ref{sec:sliced_and_projected_data} for further details)}. We leave the impact of the projection to future work.

    \red{For steep spectra (large $\beta$), $\Bpow$ is large and so $\ESarg$ becomes a worse estimate -- consistent with the \emph{signature function} literature (see, Appendix \ref{app:signature_function_relations}).}
\end{itemize}

Since a $\beta$ dependent $b$ argument seems impractical for analysis of real observations, from here on we focus on the case of a constant $b$, and explore \emph{which} value of $b$ is most appropriate under some example circumstances along with the amplitude biases induced and how to correct for them.

\subsection{Analytical Solutions of Model Correlation Functions}
\label{sec:analytical_solutions_of_model_correlation_functions}

Unbounded pure power law spectra are not physical, and regions of pure power law do not even necessarily indicate turbulent behaviour. In addition, we are also left with ambiguity in a constant $b$. We therefore consider a more nuanced approach where we attempt to determine appropriate choices of $b$ by focusing on the location of the peaks of $\Earg$ and $\ESarg$ by modelling spectra that are \emph{not} pure-power law functions. The peaks of the associated spectra are at the wavenumbers
\begin{subequations}
    \begin{align}
        \label{eqn:kp}
        \kp &= \underset{k}{\text{argmax}} \{ \Earg \},\\
        \label{eqn:kep}
        \kep &= \underset{\ke}{\text{argmax}} \{ \ESarg \},
    \end{align}
\end{subequations}
where $\kp$ and $\kep$ can be thought of as an effective energy-containing scale characteristic of the turbulence energy-containing/outer scales (\eg{} injection scale, correlation scale, \etc{})\cite{Pope00}.

From \autoref{sec:pure_powerlaw_bias} we already know the pure power-law behaviour is well captured by $\ESarg$. However, if we wish to accurately approximate the PSD using the equivalent SF, it should also accurately describe an energy-containing scale $L$. We investigate model spectra with a distinctive scale range in $\Earg$ associated with the $L \sim 1/\kp$ that transitions to a pure power law regime at $k \gg 1/L$.

We will use the following model,
\begin{align}
    \barR = D e^{-\ell/L} ,
\end{align}
where $L$ corresponds to some energy-containing scale (in this case, the correlation length). This is a case already present in the literature for Hurst parameter $H=1/2$, but we have extended to arbitrary dimensions $D$ \cite{Pope00}.

The corresponding angle-integrated spectrum is \citep{Bracewell86, Stutzki.etal98}
\begin{subequations}
    \begin{align}
        \label{eqn:integrated_spectrum_from_acf}
        \Earg &= \frac{\Omega_D^2}{(2\pi)^D} \int_0^\infty \barR \mathcal{T}_D(k \ell) \bracket{k \ell}^{D-1} \d \ell, \\
        \label{eqn:model_spec_G14_1}
        &= \frac{2}{\sqrt{\pi}} \frac{D L^D k^{D-1}}{\bracket{L^2 k^2 + 1}^{(D+1)/2}} \dfrac{\Gamma\bracket{\frac{D+1}{2}}}{\Gamma\bracket{\frac{D}{2}}} ,\\
        &= \begin{cases}
            \frac{2}{\pi} \frac{L}{1 + L^2 k^2} , & D = 1\\
            \frac{2 L^2 k}{\bracket{1 + L^2 k^2}^{3/2}} , & D = 2\\
            \frac{12}{\pi} \frac{L^3 k^2}{\bracket{1 + L^2 k^2}^{2}} , & D = 3\\
            \dots , & \\
        \end{cases}.
    \end{align}
\end{subequations}

Using \autoref{eqn:sf_variance_acf} with \autoref{eqn:equivalent_spectrum}, the equivalent spectrum is
\begin{subequations}
    \begin{align}
        \label{eqn:equivalent_spectrum_via_acf}
        \ESarg &= - \frac{\ell^2}{b} \frac{\d \barR}{\d \ell},\\
        \label{eqn:model_spec_G14_1_est}
        &= b \frac{D}{L \ke^2} e^{- \frac{b}{L \ke}} .
    \end{align}
\end{subequations}

\autoref{eqn:model_spec_G14_1} and \autoref{eqn:model_spec_G14_1_est} have local power law behaviour as
\begin{align}
    \Delta \beta(k) &= -\frac{\d \ln(\mathcal{E}_D(k))}{\d \ln(k)} = -\frac{D - 1 - 2L^2 k^2}{1 + L^2 k^2},\\
    \betaestarg &= -\frac{\d \ln( \ESarg)}{\d \ln(\ke)} = -\frac{b}{L \ke} + 2 .
\end{align}
$\Earg$ and $ \ESarg$ follow the same power law slope for $kL, \ke L \gg 1$ \ie{}$\Earg \sim k^{-2},  \ESarg \sim k_{\mathrm{e}}^{-2}$. See Appendix \ref{app:additional_analytics} for an analytical case where $ \widetilde{\mathcal{E}}_{D=1}^{S}(k_{\mathrm{e}})$ has the correct power law behaviour for general power law $\beta$.

The effective energy-containing scales ($\kp, \kep$) of \autoref{eqn:model_spec_G14_1} and \autoref{eqn:model_spec_G14_1_est} are found using $\Delta \beta(\kp) = \Delta \tilde{\beta}(\kep) = 0$, which are
\begin{align}
    \kp &= \frac{\sqrt{2D - 2}}{2 L} = \begin{cases}
        0 , & D = 1\\
        \frac{1}{\sqrt{2}} \frac{1}{L} , & D = 2\\
        \frac{1}{L} , & D = 3\\
        \dots \\
    \end{cases},\\
    \kep &= \frac{b}{2L}.
\end{align}
This naturally provides the relationship $b = \sqrt{2D - 2}$ for $D > 1$ which ensures that the peaks of \autoref{eqn:model_spec_G14_1} and \autoref{eqn:model_spec_G14_1_est} are aligned. This does not imply the amplitudes at the peaks are equal \ie{}$\E(\kp) \ne \sES(\kep)$. We merely choose a $b$ so that the peaks of the respective spectra are located at the same wavenumber ($\kp = \kep$). Note, the case for $D=1$ provides $b=0$ because the peak of \autoref{eqn:model_spec_G14_1} is at $k=0$.

\begin{sidewaystable*}
\centering
\begin{tabular}{ c c c | c c | c }
    \hline
    \hline
    $\dfrac{\overline{R}_{D}(\ell)}{A}$ & $\dfrac{\mathcal{E}_{D}(k)}{A}$ & $\dfrac{ \widetilde{\mathcal{E}}_{D}^{S}(k_{\mathrm{e}})}{A}$ & $k_{\mathrm{p}}$ & $\dfrac{k_{e,\mathrm{p}}}{b}$ & $b \sqbracket{\equiv \dfrac{b k_{\mathrm{p}}}{k_{\mathrm{e},\mathrm{p}}}}$ \\ 
    \hline
    \hline
     $D e^{-\ell/L}$ & $ 2 D L^D k^{D-1} \bracket{L^2 k^2 + 1}^{-(D+1)/2} \dfrac{\Gamma\bracket{\frac{D+1}{2}}}{\sqrt{\pi} \Gamma\bracket{\frac{D}{2}}} $ & $D b \dfrac{e^{-b/Lk_{\mathrm{e}}}}{L k_{\mathrm{e}}^2}$ & $\dfrac{\sqrt{2D - 2}}{2 L}$  & $\dfrac{1}{2L}$ & $\sqrt{2D - 2}$\\  
    $\dfrac{\bracket{DL - \ell} e^{-\ell/L}}{L}$ & $4 L^{D+2} k^{D+1} \bracket{L^2 k^2 + 1}^{-(D+3)/2} \dfrac{\Gamma\bracket{\frac{D+3}{2}}}{\sqrt{\pi}\Gamma\bracket{\frac{D}{2}}}$ & $b \bracket{DL + L - \dfrac{b}{k_{\mathrm{e}}}} \dfrac{e^{-b/Lk_{\mathrm{e}}}}{L^2 k_{\mathrm{e}}^2}$ & $\dfrac{\sqrt{2D + 2}}{2L}$ & $\dfrac{\bracket{D \pm \sqrt{D^2 + 8} + 4}}{4 L \bracket{D + 1}}$ & $\dfrac{2 \sqrt{2} \bracket{D + 1}^{3/2}}{D + \sqrt{D^2 + 8} + 4}$ \\
    $\dfrac{\bracket{D L^2 - 2 \ell^2} e^{-\ell^2/L^2}}{L^2}$ & $\dfrac{2^{-D}}{\Gamma\bracket{\frac{D}{2}}} L^{D+2}k^{D+1} e^{-L^2 k^2 / 4}$ & $2 b^2 \bracket{DL^2 + 2 L^2 - \frac{2 b^2}{k_{\mathrm{e}}^2}} \dfrac{e^{-b^2/L^2 k_{\mathrm{e}}^2}}{L^4 k_{\mathrm{e}}^3}$ & $\dfrac{\sqrt{2D + 2}}{L}$ & $ \dfrac{\sqrt{3D \pm 3 \sqrt{D^2 + 2D + 25} + 21}}{3 L \sqrt{D + 2}}$ & $\dfrac{\sqrt{6D^2 + 18D + 12}}{\sqrt{D + \sqrt{D^2 + 2D + 25} + 7}}$
\end{tabular}
\caption{\textbf{Analytical solutions for spectra using different model autocorrelation functions}. Shown is: the angle-integrated spectrum $\Earg$ and $ \ESarg$ using these different forms of the autocorrelation function $\overline{R}_{D}(\ell)$. We also show the turn-overs/peaks of the respective spectra $k_\mathrm{p}$, $k_{e,\mathrm{p}}$ and the corresponding factor $b$ required to equate $k_{e,\mathrm{p}}$ to $k_\mathrm{p}$. Note, for solutions in $k_{e,\mathrm{p}}$ with a $\pm$, we take the $+$ form in $b$ as these are more in line with alternative solutions to $b$ in \autoref{sec:filter_function}. Note that for $D=3$, the last column is equivalent to Townsend’s model eddy of a swirling fluid of a characteristic size $L$ \cite{Townsend76, Davidson15}.}
\label{tab:b_factors}
\end{sidewaystable*}

Unfortunately, this method of estimating an appropriate \emph{constant} $b$ is dependent on the underlying model of $\barR$. \autoref{tab:b_factors} shows some analytical expressions using different forms of $\barR$. Note, the third row in \autoref{tab:b_factors} is analogous to the model eddy of a swirling fluid of characteristic size $L$ for $D=3$ \red{(Townsend’s model)} \cite{Townsend76, Davidson15}. The model correlation functions and spectra in \autoref{tab:b_factors} may serve as adequate models to verify some spectral estimate behaviours. \red{We also note that some of the models in \autoref{tab:b_factors} result in non-positive $\ESarg$ for a range of $\ke \ll 1/L$. In these cases, $\ESarg$ would not be an appropriate approximation of $\Earg$ within that negative region (since $\Earg$ is a non-negative function \ie{}$\Earg \ge 0$).} \red{Similar analysis have also been to applied to the \emph{signature functions} (for $D=3$) to examine their validity for representing the turbulence energy-containing range \cite{Davidson15} (see, Appendix \ref{app:signature_function_relations} for more details).}

It has been noted in previous work that local features in the PSD are non-local in the SF (see \autoref{eqn:spectral_representation}) and the transfer of these local features from one space to the other is dependent on the functional form of the PSD \cite{Donzis.Sreenivasan10, Davidson15}. It would appear that the equivalent spectrum inherits this property -- which we discuss in the next section. Consequently, when the constant $b$ equates the energy-containing scales, its value depends on the functional form of the PSD. Therefore, \emph{there is no universal constant $b$}.

\subsection{$\ESarg$ as a Filtered Function of $\Earg$}
\label{sec:filter_function}
An alternative method we can use to estimate a \emph{constant} $b$ is to describe \autoref{eqn:equivalent_spectrum} as a filter on $\Earg$
\begin{align}
    \label{eqn:equiv_spectrum_filter}
    \ES\bracket{\frac{b}{\ell}} = \int_0^\infty \Earg \mathcal{G}_{D}(k\ell) \d k ,
\end{align}
where
\begin{align}
    \nonumber
    \mathcal{G}_{D}(k \ell) &= \frac{1}{b}\ell^2 \frac{\partial}{\partial \ell}\sqbracket{1 - \mathcal{T}_{D}(k\ell)},\\
    \label{eqn:filter_func}
    &= 2^{D/2 - 1} \Gamma\bracket{D/2} \frac{\ell}{b} \bracket{k \ell}^{2-D/2} \mathcal{J}_{D/2}(k\ell).
\end{align}
which is calculated by differentiating under the integral of
\begin{align}
    \ES\bracket{\frac{b}{\ell}} = \frac{1}{b} \ell^2 \frac{\partial}{\partial \ell}\sqbracket{\int_0^\infty \sqbracket{1 - \mathcal{T}(k\ell)} \mathcal{E}_{D}(k) \d k},
\end{align}
when the spectral representation of the SF (\autoref{eqn:spectral_representation}) is used in \autoref{eqn:equivalent_spectrum}.

Ideally, $\mathcal{G}_{D}(k\ell)$ would be a $\delta$ function-like to select from $\Earg$ the appropriate energies for the corresponding physical scale $\ell$. We now show that \autoref{eqn:filter_func} is \emph{not} $\delta$ function-like. 

\begin{figure}
    \centering
    \includegraphics[scale=1]{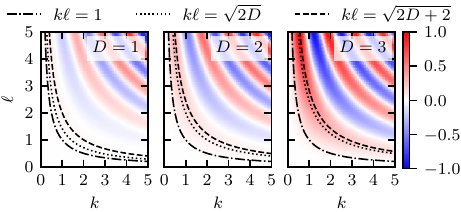}
    \caption{Filter function $\mathcal{G}_{D}(k\ell)$ normalized by its maximum (in the generated grid) to [-1, 1] for $D=1,2,3$. The black dash-dotted, dotted, and dashed contours correspond to $k \ell=1$, $k \ell=\sqrt{2D}$, and $k \ell=\sqrt{2D+2}$ respectively.}%
    \label{fig:filter_func}%
\end{figure}

To associate $k \ell$ with $b$ we say that $k \ell = b \mathcal{K}$ where $\mathcal{K} = k/\ke$ and assert $\mathcal{K} = 1$ \ie{}$\ke \equiv k$. \autoref{fig:filter_func} shows \autoref{eqn:filter_func} as a function of $k$ and $\ell$ along with contours corresponding to $b=1$, $b=\sqrt{2D}$, and $b=\sqrt{2D + 2}$. We have chosen these values based on extensions of literature results \cite{Davidson.Pearson05, Davidson15} to $D=1,2$, along with low-order coefficients of Taylor expansions of $\mathcal{G}_{D}(k\ell)$ about $k\ell=0$. The first peak of $\mathcal{G}_{D}(k\ell)$ is better represented by $\sqrt{2D}$ to $\sqrt{2D + 2}$. If we crudely assume that $\mathcal{G}_{D}(k\ell)$ behaves $\delta$ function-like, then $\mathcal{G}_{D}(k\ell)$ preferentially selects from $\Earg$ at the locations corresponding to $\sqrt{2D}$ to $\sqrt{2D + 2}$.

\begin{figure}
    \centering
    \includegraphics[scale=1]{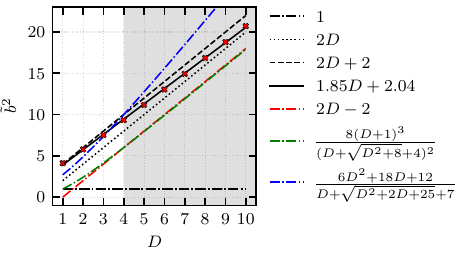}
    \caption{The location of the first peaks ($\tilde{b} = x_{\text{peak}} = \text{argmax} \{ \widetilde{\mathcal{G}}_{D}\bracket{x} \}$) of the filter $\widetilde{\mathcal{G}}_{D}(x) \sim x^{2-D/2} \mathcal{J}_{D/2}(x)$ as a function of dimension $D$. The first peak locations (red crosses) are estimated using numerical methods and the black solid line is the linear fit. The dashed and dotted black lines are some reasonable approximations. The red, green, and blue dash-dotted lines correspond to the estimates of $b$ from \autoref{tab:b_factors}. Note, the estimates from \autoref{tab:b_factors} correspond to the \emph{integrated} influence of $\mathcal{G}_{D}(k\ell)$ with $\mathcal{E}_{D}(k)$ rather than just one part of the integrand.}%
    \label{fig:filter_b_peak}%
\end{figure}

As a first-order approximation, assuming the peak of \autoref{eqn:filter_func} is where the peak of the corresponding $\ESarg$ is located could provide a suitable method for estimating an appropriate $b$ for an unknown $\Earg$. To find where these peak contours of the $\mathcal{G}_{D}(k \ell)$ lie, we first remove the extra $\ell$ dependence, and define the following 1D function
\begin{align}
    \label{eqn:alternative_filter}
    \widetilde{\mathcal{G}}_{D}\bracket{x=k\ell} \equiv \mathcal{G}_{D}\bracket{k \ell}\frac{b}{\ell} \sim x^{2-D/2} \mathcal{J}_{D/2}(x).
\end{align}
We numerically evaluate $\tilde{b} = x_{\text{peak}} = \text{argmax} \{ \widetilde{\mathcal{G}}_{D}\bracket{x} \}$. As $\ell$ linearly scales the filter for a fixed $k$, $\tilde{\mathcal{G}}_{D}\bracket{x}$ is an appropriate substitute for $\mathcal{G}_{D}(k\ell)$. We show $\tilde{b}^2$ as a function of $D$ in \autoref{fig:filter_b_peak}. The linear fit $\tilde{b}^2 = 1.85 D + 2.04$ is approximately our stated $\tilde{b}^2 = 2D + 2$ contour in \autoref{fig:filter_func} for $D < 4$. Estimates from \autoref{tab:b_factors} are also included in \autoref{fig:filter_b_peak}. These estimates are obtained from the peaks of \autoref{eqn:equiv_spectrum_filter} \ie{}rather than from just one part of the integrand.

As $k\ell$ increases, $\mathcal{G}_{D}(k\ell)$ oscillates about 0 (see, \autoref{fig:filter_func}) which shows that $\mathcal{G}_{D}(k\ell)$ is not $\delta$ function-like. The oscillations cause $ \ESarg$ to be influenced by $\Earg$ across \emph{all} scales \ie{}the relation has non-local influences. It also explains the dependence of $\mathcal{B},b$ on the form of $\Earg$ as seen in this, and the previous sections. For a fixed $\ell$, the filter peaks at $k \ell \sim \sqrt{1.85D + 2.04}$ and oscillates around 0 with increasing $k$. As a crude approximation, we could say $\mathcal{G}_{D}(k \ell)$ is $\delta$ function-like and the contributions (oscillations) at large $k$ cancel out. However, for a PSD with large $\beta$, the difference in PSD amplitude between adjacent wavenumbers is larger, meaning successive contributions from $\Earg$ with $\mathcal{G}_{D}(k\ell)$ in \autoref{eqn:equiv_spectrum_filter} would \emph{not} perfectly cancel out. For $D=1,2$, the oscillations grow larger in amplitude with increasing $k$ meaning the successive contributions would also not perfectly cancel out. For small $\beta$, the amplitude between adjacent wavenumbers is similar, meaning successive contributions would approximately cancel, resulting in a net zero contribution for scales $k \ell \gg 1$. The influence of the non-cancelled non-local contributions propagates into $b$ and $\mathcal{B}$ and explains the observation from \autoref{fig:analytical_amplitude_bias} that for all $D=1,2,3$ at large $\beta$, the bias is large -- and also explains the dependence on the form of the PSD. For $D=1$, the bias is larger than $D=2,\,3$ because the growth in amplitude of the oscillations is greater than $D=2,\,3$.

We conclude this section by suggesting
\begin{align}
    \label{eqn:bp_est}
    \begin{split}
        \bpest &= \begin{cases}
            \sqrt{2D-2}\,&\text{for }D>1\\
            1\,&\text{for }D=1
        \end{cases}
        = \begin{cases}
            1\,&\text{for $D=1$}\\
            \sqrt{2}\,&\text{for $D=2$}\\
            2\,&\text{for $D=3$}\\
            \dots
        \end{cases},
    \end{split}
\end{align}
as a $D$-dependent $b$ factor. The slope of $(b^{\mathrm{est}}_{\mathrm{p},D})^2$ with $D$ is approximately the slope of the linear fit shown in \autoref{fig:filter_b_peak}. \autoref{eqn:bp_est} also closely represents the first two analytical solutions in \autoref{tab:b_factors} without the non-physical $b=0$ for $D=1$ from the $b=\sqrt{2D-2}$ solution. The subscript-$p$ represents that this $b$ is associated with correcting the \emph{peak} (the energy-containing scale) of the equivalent spectrum. This is in contrast to the pure power law regime where the $b$ to equate (all) the amplitudes is analytically known ($b^{\mathrm{pow}}_{D}$, \autoref{eqn:analytical_b_for_powerlaw}). It is the case that $\bpest \ne b^{\mathrm{pow}}_{D}$. Therefore, aligning the energy-containing scales (using $b=\bpest$) comes at the cost of the amplitude bias in the power law regime (see \autoref{fig:numerical_B_powerlaw_estimate}). The accuracy of $\bpest$ to align the energy-containing scales is discussed further in the next section. In \autoref{sec:powerlaw_approx}, we show that the cost of using $\bpest$ (instead of $b^{\mathrm{pow}}_{D}$) in the pure power law regime can be corrected for.

\subsection{Model Spectrum with an energy-containing Scale}
\label{sec:exp_analytical_bias}
Now, we define a model angle-integrated spectrum,
\begin{align}
    \label{eqn:exp_spectrum}
    \Earg = \Omega_{D} k^{D-1} \sqbracket{\bracket{\frac{k}{k_0}}^{-\alpha} e^{-k_0^2/k^2}},
\end{align}
that contains exponential growth at small $k$ which represents a turbulence energy-containing range that has a well-defined effective energy-containing scale (peak) at $\kp = \sqrt{2k_0^2/\beta}$. The model spectrum then transitions to pure power law decay $\beta = \alpha - D + 1$ at $k \gg \kp$.

The corresponding amplitude bias function is,
\begin{align}
    \label{eqn:meijerG_bias}
    B^{\mathrm{exp}}_{D}\bracket{\frac{k}{k_{\mathrm{p}}}, \beta, b} &= 2^{1-\beta} b^{\beta - 1} \Gamma\bracket{\frac{D}{2}}\times\\
    \nonumber
    &\mathcal{M}_{3,0}^{0,2}\bracket{\frac{\beta-1}{2},\,1\,\,\,\frac{\alpha}{2} \bigg|\frac{8}{b^2 \beta} \frac{k^2}{k_{\mathrm{p}}^2}} e^{\frac{\beta}{2}\frac{k_{\mathrm{p}}^2}{k^2}},
\end{align}
where $\mathcal{M}^{m,n}_{p,q}$ is the Meijer G-function \cite{Beals.Szmigielski13}. The Meijer G-function is unfortunately more analytically complicated than even the generalized hypergeometric functions! Although they can be analytically difficult to analyze, we use numerical methods that are readily available through Python packages such as \href{http://www.mpmath.org}{\texttt{mpmath}}. When \autoref{eqn:exp_spectrum} is the Fourier spectrum, the equivalent spectrum has a systematic bias described by \autoref{eqn:meijerG_bias}. This bias is wavenumber dependent, unlike the pure power-law example from \autoref{sec:pure_powerlaw_bias}. Note that \autoref{eqn:meijerG_bias} is applicable to \textit{only} \autoref{eqn:exp_spectrum} and under different $\Earg$ (\eg{}different powers in the exponential term) the corresponding bias function will be different for the reasons described in \autoref{sec:filter_function}. 

\begin{figure*}
    \centering
    \includegraphics[scale=1]{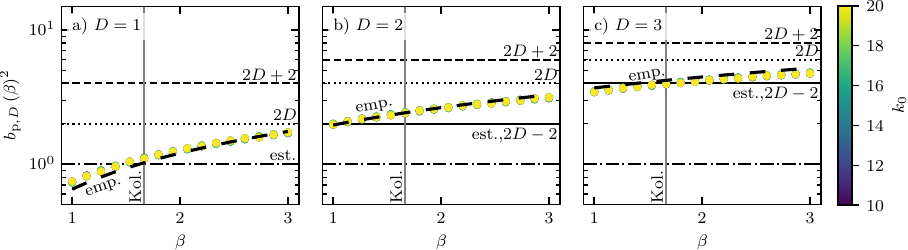}
    \caption{$b$ factor ($b_{\mathrm{p}}$) corresponding to aligning the position of the peaks of $\widetilde{\mathcal{E}}_{D}^{S}(k_{\mathrm{e}})$ with $\mathcal{E}_{D}(k)$ using the model spectrum \autoref{eqn:exp_spectrum} for different $\beta$ and $k_0$ values (circles). Note, the circle markers are coloured representing different $k_0$ (shown by the colourbar), but $\bp$ appears almost independent of $k_0$. The black horizontal lines correspond to $1$ (dash-dot), $2D-2$ (solid), $2D$ (dotted), and $2D+2$ (dashed) as discussed in \autoref{sec:filter_function} with $\bpest$ denoted as ``est.'' where appropriate. The $\beta$-dependent thick black dashed line is the empirical formula \autoref{eqn:empirical_b_estimate} ($\bpemp$, denoted as ``emp.''). The solid vertical gray line represents the (Kolmogorov) power law slope of $-5/3$.}%
    \label{fig:semi_analytical_peak_b}%
\end{figure*}

\autoref{fig:semi_analytical_peak_b} shows the $b$ required to align the position of the peaks of $\ESarg$ with $\Earg$. We analytically solve for $ \ESarg$ using \autoref{eqn:equiv_spectrum_filter} with \autoref{eqn:exp_spectrum}\footnote{
\begin{align}
    \sESarg = 2^{2-\beta} \pi^{D/2}  k_0^{\alpha} b^{\beta - 1} \ke^{-\beta} \mathcal{M}_{3,0}^{0,2}\bracket{\frac{\beta-1}{2},\,1\,\,\,\frac{\alpha}{2} \bigg|\frac{4}{b^2} \frac{k^2}{k_{0}^2}}
\end{align}}
then numerically find $\kep$.

We see the complexity initially discussed in \autoref{sec:filter_function}: whilst the filter preferentially selects at the scale associated with $b \approx \sqrt{1.85D + 2.04}$, it is the \emph{integration} with (the nonlinear) $\mathcal{E}_{D}(k)$ that determines the actual peak location of $ \ESarg$. The factor $\bp$ naturally depends on the form of the $\Earg$. The previous fit of $\bp^2 = 1.85D + 2.04$ significantly differs from the analytical solutions. Given \autoref{eqn:exp_spectrum}, \autoref{fig:semi_analytical_peak_b} shows $b$ depends on $\beta$ and $D$ with negligible dependence on $k_0$. At (Kolmogorov) $\beta \sim 5/3$, the choice of $\bp = \bpest$ (\autoref{eqn:bp_est}) approximates the analytical $\bp$ (particularly for $D=1,3$). We also provide a $\beta$ and $D$ dependent \emph{empirical} estimate by fitting to the data shown in \autoref{fig:semi_analytical_peak_b}
\begin{align}
    \label{eqn:empirical_b_estimate}
    \bpemp(\beta) = \sqrt{\frac{\beta + D - 1}{2}} + \frac{3(D-1)+1}{10}.
\end{align}
that approximates the numerical $\bp$ across $\beta$ values. However, practically, using $b=\bpemp$ would require \emph{a priori} knowledge of $\beta$, and therefore, we largely ignore this estimate.

\begin{figure}
    \centering
    \includegraphics[scale=1]{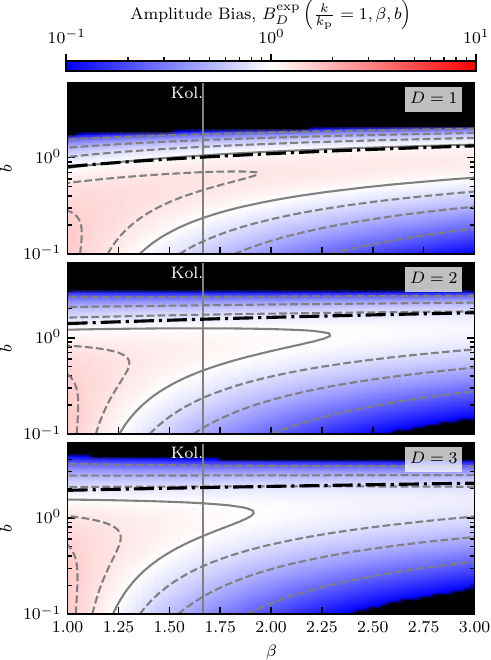}
    \caption{Similar to \autoref{fig:numerical_B_powerlaw_estimate}. The bias (\autoref{eqn:meijerG_bias}) at the peak of the spectrum $k = \kp$ for an angle-integrated spectrum of the form \autoref{eqn:exp_spectrum}, for Euclidean dimensions $D=1$, $D=2$, and $D=3$. The line contours correspond to linear spacings of 0.25 starting from $\Bexp=0.25$ to $1.75$ with $\Bexp = 1$ represented by the solid gray line. The solid vertical gray line represents the (Kolmogorov) power law slope of $5/3$. The dash-dotted black line corresponds to $\bpemp$ (\autoref{eqn:empirical_b_estimate}).}%
    \label{fig:numerical_B_estimate}%
\end{figure}

\autoref{fig:numerical_B_estimate} shows \autoref{eqn:meijerG_bias} at the (equivalent) wavenumber corresponding to the effective energy-containing scale of \autoref{eqn:exp_spectrum}. We note the following:
\begin{itemize}
    \item There is no $b$ that allows $\Bexp\bracket{\frac{k}{\kp} = 1, \beta, b} = 1$ for all $\beta \in [1, 3]$ for $D=2,3$. So even for a $\beta$ dependent $b$, there will still be a systematic bias in the equivalent spectrum.
    \item For $D=1$, $\bpemp$ closely follows the $\Bexp=1$ contour line: the peaks of $\Earg$ and $\ESarg$ are aligned in wavenumber ($\kp \approx \kep$) and amplitude: $\E(\kp) \approx \ES(\kep)$.
    \item For $D=2$ and $D=3$, $\bpemp$ closely follows a constant contour $\Bexp \approx 0.75$. The peaks are aligned in wavenumber and differ in amplitude for all $\beta$ by the same constant factor: $\E(\kp) \sim \ES(\kep)$.
\end{itemize}


\red{We note that at scales $\ke \ll \kp$, $\ESarg$ becomes negative and therefore, would not physically represent $\Earg$ at those scales.} However, in this section, we have determined appropriate $b$ values ($\bemp$ and $\approx \best$) that align the effective energy-containing scales of $\ESarg$ to those of $\Earg$. We will investigate the power law behaviour in the next section and show that $\Bexp \rightarrow \Bpow$ as $\ke \gg \kp$.



\subsection{The Power Law Limit}
\label{sec:powerlaw_approx}

Investigation of \autoref{eqn:meijerG_bias} indicates that $\Bexp \rightarrow \mathrm{const}$ for a fixed $\beta$ and $b$ as $k \gg \kp$. 

\begin{figure}
    \centering
    \includegraphics[scale=1]{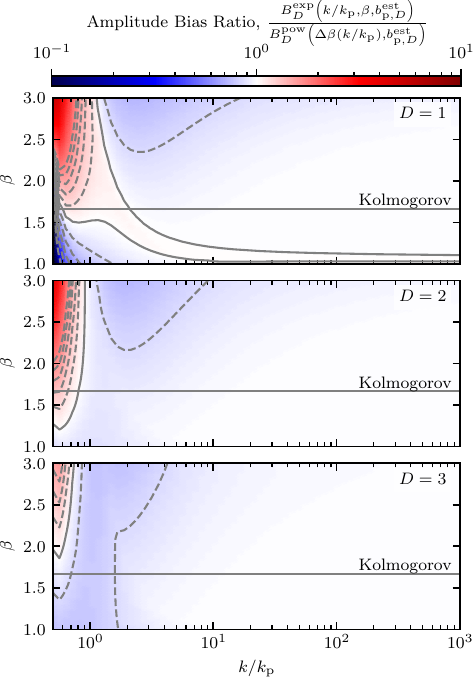}
    \caption{The ratio of the amplitude biases \autoref{eqn:meijerG_bias} and \autoref{eqn:powerlaw_bias} as a function of $k/k_{\mathrm{p}}$ and $\beta$, for Euclidean dimensions $D=1$, $D=2$, and $D=3$. The line contours correspond to linear spacings of 0.25 starting from $\Bexp=0.25$ to $1.75$ with $1$ represented by the solid gray line. The solid horizontal gray line represents the (Kolmogorov) power law slope of $5/3$.}%
    \label{fig:numerical_B_estimate_k_vs_powerlaw_2}%
\end{figure}

We have analytically calculated the local power law index
\begin{align}
    \label{eqn:local_powerlaw_est}
    \Delta \tilde{\beta}(\ke) = -\frac{\d \ln(\ESarg)}{\d \ln(\ke)},
\end{align}
of $\ESarg$ corresponding to \autoref{eqn:exp_spectrum}. We note that we need to clamp $\betaest \in (1, 3)$ as required by \autoref{eqn:powerlaw_bias}. We provide $\betaest(\ke)$ as a $\ke$-dependent argument to $\Bpow$ and compare $\Bpow(\betaest(k/\kp), b)$ to $\Bexp(k/\kp, \beta, b)$. \autoref{fig:numerical_B_estimate_k_vs_powerlaw_2} shows that we have a biased spectrum at $k \sim \kp$, and a unbiased spectrum at $k \gtrsim 10 \kp$ as the ratio clearly asymptotes to 1. Therefore, we can acknowledge that $\Bexp \rightarrow \Bpow$ at large $k$. This is consistent with the findings in \autoref{sec:analytical_solutions_of_model_correlation_functions} and Appendix \ref{app:additional_analytics}: $\ESarg$ will exhibit the correct pure power law behaviour (for reasonable underlying $\Earg$) with an intrinsic amplitude bias as described by $\Bpow$.

Previous works\cite{Chhiber.etal18, Thepthong.etal23} have focused on the turbulence inertial range and have made the approximation $\overline{S}_{D}(\ell) \propto \ell^{\eta}$ where $\eta = \beta - 1$ and therefore, effectively \autoref{eqn:equivalent_spectrum} is of the form\footnote{Other literature\cite{Thepthong.etal23} has a factor of $1/4$ (instead of our $1/2$) because they have normalized to the \emph{one-sided} $\Earg$ \ie{}
\begin{align}
    \frac{\avg{s_{D}(\vec{x})^2}}{2} = \int_0^\infty \Earg \d k,
\end{align}
alternatively, thought of as normalizing such that $\Earg \d k$ represents the energy-per-unit mass. When we use \autoref{eqn:sf_variance} and \autoref{eqn:sf_fundamental_theorem}, which states
\begin{align}
    \avg{s_{D}(\vec{x})^2} = \int_0^\infty \Earg \d k.
\end{align}
which means that $\Earg \d k$ represents twice the energy-per-unit mass.}
\begin{align}
    \label{eqn:powerlaw_ESF_approx}
    \widetilde{\mathcal{P}}^{S}_{D}(\ke) \equiv \frac{1}{2} \frac{\eta}{b} \ell \barS \sim \mathcal{B} \ESarg,
\end{align}
where $\widetilde{\mathcal{P}}^{S}_{D}(\ke)$ is the power law region approximation to $\ESarg$. For an unbounded pure power law spectrum (\autoref{eqn:powerlaw_spectrum}) then the approximation $\barS \sim \ell^{\eta}$ is exact and therefore \autoref{eqn:powerlaw_ESF_approx}, \autoref{eqn:analytical_equivalent}, and \autoref{eqn:powerlaw_bias} are exact. Note that the $b$ cases discussed in \autoref{sec:pure_powerlaw_bias} are still present -- either $b$ is $\beta$ dependent, or $b$ is constant with the specific value being ambiguous (see previous sections) and $\widetilde{\mathcal{P}}^{S}_{D}(\ke)$ has a systematic amplitude bias described by \autoref{eqn:powerlaw_bias}.

\begin{figure}
    \centering
    \includegraphics[scale=1]{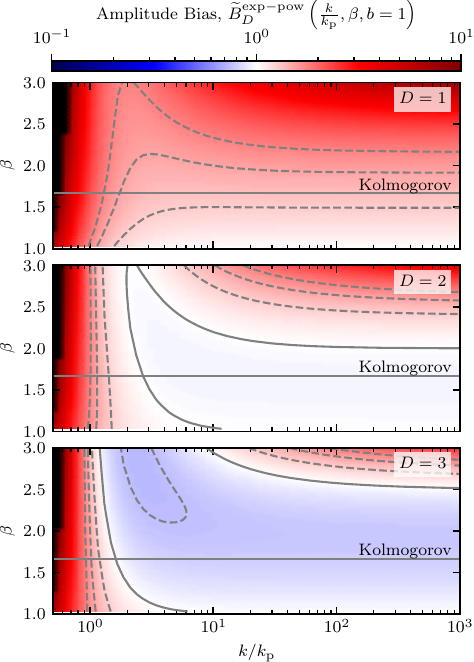}
    \caption{The amplitude bias (\autoref{eqn:pow_approximation_exp_bias}) of the power law approximation for the model spectrum \autoref{eqn:exp_spectrum} as a function of $k/\kp$ and $\beta$ with $b=1$, for Euclidean dimensions $D=1$, $D=2$, and $D=3$. The line contours correspond to linear spacings of 0.25 starting from $\widetilde{B}_{D}^{\mathrm{exp-pow}}=0.25$ to $\widetilde{B}_{D}^{\mathrm{exp-pow}}=1.75$ with 1 represented by the solid gray line. The solid horizontal gray line represents the (Kolmogorov) power law slope of $5/3$.}%
    \label{fig:numerical_B_estimate_k_vs_powerlaw_2_pow_est}%
\end{figure}

Using the model spectrum of \autoref{eqn:exp_spectrum} with the equivalent spectrum power law approximation (\autoref{eqn:powerlaw_ESF_approx}), we get the following amplitude bias function
\begin{align}
    \label{eqn:pow_approximation_exp_bias}
    \begin{split}
        \widetilde{B}_{D}^{\mathrm{exp-pow}}&\bracket{\frac{k}{\kp}, \beta, b} = \frac{F}{b} e^{\frac{\beta}{2} \frac{\kp^2}{k^2}} \bracket{\frac{\kp}{k}} \Bigg[\\
        &2^{3\beta/2} D b \sqrt{\beta} \bracket{\frac{k}{\kp}}^\beta \Gamma\bracket{\frac{\beta-1}{2}} \\
        - 4 \sqrt{2}b^\beta &\beta^{\beta/2} \frac{k}{\kp} \mathcal{M}_{3,0}^{0,2}\bracket{ \frac{\beta+1}{2},1\,\, \frac{\alpha}{2} \bigg| \frac{8}{b^2 \beta} \frac{k^2}{\kp^2}} \Gamma\bracket{\frac{D}{2} + 1} \Bigg]
    \end{split}
\end{align}
where $F = 2^{-\beta-3/2} \beta^{-\beta/2} \bracket{\beta - 1}/D$. This bias function is shown in \autoref{fig:numerical_B_estimate_k_vs_powerlaw_2_pow_est} for $b=1$. In the pure power law region $k \gg \kp$, the bias asymptotes to a constant for a fixed $\beta$. As with $\Bexp$ (\autoref{eqn:meijerG_bias}), $\widetilde{B}_{D}^{\mathrm{exp-pow}}$ approaches $\Bpow$ for $k \gg \kp$.

Previous applications of $\ESarg$ in the literature have used the pure power law approximation form $\widetilde{\mathcal{P}}^{S}_{D}(\ke)$ (\autoref{eqn:powerlaw_ESF_approx}) \cite{Huang.etal09, Chhiber.etal18, Thepthong.etal23}. Similar to \autoref{fig:numerical_B_powerlaw_estimate} and \autoref{fig:numerical_B_estimate_1d}, \autoref{fig:numerical_B_estimate_k_vs_powerlaw_2_pow_est} explains its success. For any reasonable underlying spectrum $\Earg$, the turbulence inertial range is well captured by $\ESarg$ (and $\widetilde{\mathcal{P}}^{S}_{D}(\ke)$) with an amplitude bias that is described by $\Bpowarg$.

We have shown that we can debias (apply a systematic correction to) the equivalent spectrum in the pure power law regime, assuming that we can estimate the correct power law. We have also shown that the equivalent spectrum will exhibit the correct power law behaviour at large $k$ (for $1 < \beta < 3$) \cite{Monin.Iaglom75, Stutzki.etal98, Pope00}. This will form the basis for a model independent debiasing technique which is introduced and discussed in the next section.

\subsection{Summary}
\label{sec:bias_section_summary}
In summary of this section, we have found that the equivalent spectrum has a systematic bias in amplitude for different Fourier spectra. In the pure power-law case, this bias is readily determined and depends on $\beta,\,D$, and $b$. We have examined two cases. Firstly, if we assume $b$ is $\beta$-dependent, then this systematic bias can be cancelled out. However, this method would be difficult to implement on real observations. In addition, the Fourier spectrum is of a different form \eg{}contains an exponential growth region, then a $\beta$-dependent $b$ cannot exactly equate the equivalent spectrum to the Fourier spectrum at the peaks of the spectra -- there will still be a systematic bias in the equivalent spectrum. Secondly, we could assume $b$ is constant and obtain a physically motivated relationship between wavenumber and physical lag-scale $\ke = b/\ell$. Although there is ambiguity in values for $b$ (that are dependent on the form of the Fourier spectrum), we provide constant $b$ values that are appropriate to align the peaks of the equivalent spectrum to the Fourier spectrum. We can then apply a general-purpose debiasing technique to reduce the systematic bias present in the equivalent spectrum to obtain a better estimate of the Fourier spectrum. This method is far more practical for implementation on real observations.


\section{Implementation and Validation}
\label{sec:implementation_testing}

In \autoref{sec:amplitude_and_wavenumber_biases} (particularly, \autoref{sec:powerlaw_approx}), we have built the intuition that the behaviour of $\ESarg$ is consistent with $\Earg$ for regions where $\Earg$ is a power law. We will start this section by introducing the practical implementation of $\ESarg$ and the method for de-biasing based on \autoref{sec:powerlaw_approx}. This will do away with the need for the bias functions which requires modelling the underlying spectrum. We then validate the implemented method on fBm fields and compare the results to analytical expressions from \autoref{sec:amplitude_and_wavenumber_biases}.

\subsection{Implementation}
\label{sec:implementation}
To calculate our estimate $\ESarg$, via the second-order structure function $\barS$, we perform the following steps in order, with bullet points indicating recommendations and additional details:
\begin{description}
    \item[Step 1] Calculate the averaged second-order structure function $\overline{S}_{D}(\ell)$.
    \begin{itemize}
        \item Calculating $\barS$ can be done using codes like \texttt{fastSF} \citep{Sadhukhan.etal21} or our provided Python code on \href{https://github.com/mab68/Equivalent_Spectrum}{Github}.
        \item Usually, $\overline{S}_{D}(\ell)$ has low amplitude-high frequency oscillations which can propagate into the derivative(s) when calculating $\ESarg$ which results in a noisy spectral estimate \cite{Ahnert.Abel07, VanBreugel.etal20}. Therefore, we suggest binning, or smoothing $\overline{S}_{D}(\ell)$ before moving onto the next step. We urge readers to exercise caution when binning the SF. Over-binning can lead to the loss of information that reflects the true turbulence statistics -- associated with intermittency effects, for example -- and could bias scaling exponent estimates \cite{Dyrud.etal08, Bentkamp.Wilczek25}. Future work will need to examine the influence of different binning procedures and binning scales.

        In this paper, we bin $\barS$ in log-space \ie{}the bins are equally spaced in log-space. Due to our binning implementation, depending on the number of bins and the span of the data, some bins incorporate no data. In those cases, we interpolate those missing values (in log-space) based on the surrounding bins. We typically use between 16 and 64 bins. This is chosen based on visually comparing the binned to un-binned $\barS$ -- we wish to retain as many features as possible whilst removing any high-frequency oscillations (which are particularly present at large $\ell$).
        \item Practically, at large $\ell$, $\barS$ tends to have large fluctuations around the value of twice the variance (\autoref{eqn:sf_variance}) as it samples inhomogeneities beyond the correlation scale \cite{DudokdeWit04}. We recommend stopping at a maximum lag $\ell_{\mathrm{max}} \lesssim \frac{N}{2} \Delta x$ where $\Delta x = \mathcal{L}/N$ is the discretized sampling spacing.
    \end{itemize}
    \item[Step 2] Create the ``Uncorrected'' estimate $\ESarg$ by using \autoref{eqn:equivalent_spectrum} with an appropriate $b$.
    \begin{itemize}
        \item We use \href{https://numpy.org/}{\texttt{numpy.gradient}} which uses a second-order central differences method with first-order forwards/backwards differences at the boundaries.
        \item Appropriate $b$ are: $\best$ (\autoref{eqn:bp_est}), and $\bemp$ (\autoref{eqn:empirical_b_estimate}). To use $\bemp$ will require either \emph{a priori} spectral slope knowledge, or a second correction will need to be applied after an initial guess of $b$ which can be used to understand the appropriate $\beta$ and therefore $\bemp$. We use $\best$.
        \item $\ESarg$ creates an estimate for $\Earg$ that still requires the systematic correction to the amplitude (as shown in the analysis in \autoref{sec:amplitude_and_wavenumber_biases}). The ``Uncorrected'' estimate will form an initial guess and its properties can be used to better understand the correction required.
        \item We interpolate the $\ke$ of $\ESarg$ onto the same $k$ as $\Earg$. The discrete wavenumber spacing is $m \Delta k$ where $\Delta k = 2\pi / \mathcal{L}$ and the discrete lag spacing is $\ell = n \Delta x$ for integers $n,m \in [1,N/2]$. The equivalent wavenumber $\ke = z \Delta \ke$ will therefore have non-integer spacing according to $z = n^{-1}$ where $\Delta \ke = bN/\mathcal{L}$. Given $k \in [2 \pi / \mathcal{L}, \pi N / \mathcal{L}]$ for $m \in [1, N/2]$ and $\ke \in [2 b / \mathcal{L}, b N / \mathcal{L}]$ for $z^{-1} = n \in [1, N/2]$, it would be natural to conclude $b=\pi$. However, \autoref{sec:amplitude_and_wavenumber_biases} suggests otherwise. As a result, we cut $\ESarg$ so that $\mathrm{max}\{ \ke \} = \mathrm{min}\{[\pi N/\mathcal{L}, bN/\mathcal{L}]\}$ and $\mathrm{min}\{ \ke \} = \mathrm{max}\{[2\pi/\mathcal{L}, 2 b / \mathcal{L}]\}$.
    \end{itemize}
    \item[Step 3] Estimate the local power law slope of the ``Uncorrected'' estimate $\betaestarg = -\frac{\d \ln ( \widetilde{\mathcal{E}}_{D}(k_{\mathrm{e}}))}{\d \ln (k_{\mathrm{e}})}$.
    \begin{itemize}
        \item Once again, we use \href{https://numpy.org/}{\texttt{numpy.gradient}}.
        \item We emphasize that this does \emph{not} require a-priori knowledge of the spectral index - a local power law index is estimated numerically, at every $k$-value on the grid.
        \item $\betaestarg$ is (roughly proportional to) the second derivative of $\overline{S}_{D}(\ell)$ which emphasizes the need to bin and/or smooth $\overline{S}_{D}(\ell)$, $\widetilde{\mathcal{E}}_{D}^{S}(k_{\mathrm{e}})$, and $\betaestarg$. Noise from the first derivative ($\ESarg$) propagates into the second derivative ($\betaest$) \cite{Wood82}.
        \item Since $\Bpowarg$ is only valid for $1 < \beta < 3$ (\autoref{sec:pure_powerlaw_bias}), we clamp to $1.01 < \betaestarg < 2.99$.
        \item We apply a final step of binning and interpolation. We bin in log-space to between $4$ and $16$ data-points. We find that this is often enough to describe the local power law for power law spectra. Interpolate back onto $\ke$ for easy use in the next step.
    \end{itemize}
    \item[Step 4] Finally, take the ``Uncorrected'' spectrum and estimate the ``Debiased'' form of the equivalent spectrum: $\mathcal{B}^{-1}\ESarg$ where $\mathcal{B} = \Bpow(\betaestarg, b)$.
    \begin{itemize}
        \item \autoref{fig:numerical_B_estimate_k_vs_powerlaw_2} is equivalent to the error obtained by using this local de-biasing technique for a Fourier spectrum of the form \autoref{eqn:exp_spectrum}. This is suitable as a general approach method where the underlying spectrum is largely unknown, or the analytical formula for the bias is difficult to obtain. Additionally, the $\Gamma$-functions in \autoref{eqn:powerlaw_bias} are more easily tractable from Python packages like \href{https://scipy.org/}{\texttt{scipy}} compared to $\mathcal{M}^{0,2}_{3,0}$.
    \end{itemize}
\end{description}
We provide these steps as Python code on \href{https://github.com/mab68/Equivalent_Spectrum}{Github}.

\subsection{fBm Validation}
\label{sec:fbm_validation}
We now consider fractional Brownian motion fields (fBm) with spectral shape specified in \autoref{eqn:exp_spectrum}. Our fBm fields are used solely for testing purposes as we know exactly what power law we expect and they are quick to generate. We generate the field $s_{D}(\vec{x})$ using a spectral synthesis method\cite{Barnsley.etal88} by defining:
\begin{subequations}
    \begin{align}
        \hat{s}_{D}(\vec{k}) &= \sqrt{\frac{E_D(\vec{k})}{E^{\text{total}}}} \bracket{\cos(\vec{\theta}) + i \sin(\vec{\theta})},\\
        \label{eqn:fbm_pspec}
        E_D(\vec{k}) &= \begin{cases} 
          0, & \abs{\vec{k}} < k_{\mathrm{min}} \\
          0, & \abs{\vec{k}} > k_{\mathrm{max}} \\
          \bracket{\frac{\abs{\vec{k}}}{k_0}}^{-\alpha} e^{-k_0^2/\abs{\vec{k}}^2}, & \text{otherwise} \\
       \end{cases},
    \end{align}
\end{subequations}
where $E^{\text{total}} = \int_{-\infty}^{\infty} E_{D}(\vec{k}) \d^{D} \vec{k}$ is the total energy, (\ie{}we have normalized so that the energy of the fBm field is $1$), and $\vec{\theta}$ are the random phases sampled from a uniform distribution from $0$ to $2\pi$. Practically, when the number of grid points $N$ is uniformly sampled in the physical domain $\mathcal{L}$: $k_{\mathrm{min}}$ corresponds to the discrete wavenumber spacing $\Delta k = 2\pi/\mathcal{L}$ and $k_{\mathrm{max}}$ corresponds to the Nyquist wavelength $\pi N/\mathcal{L}$. We then take the inverse Fourier transform of $\hat{s}_D(\vec{k})$ to obtain the fBm field $s_D(\vec{x})$.

Fractional Brownian motion fields have been used extensively to validate methods for interstellar turbulence \cite{Brunt.Heyer02, Miville-Deschenes.etal03, Esquivel.etal03, Ossenkopf.etal06, Chepurnov.Lazarian09, Roman-Duval.etal11}, with modifications using Perlin noise as the basis functions or exponentiating \citep{Panopoulou.etal17, Bates.etal20}. They have similarly been used for the ICM, when measuring the turbulence structure from rotation measure observations \citep{Vogt.Ensslin05} and ICM telescope \& method testing \citep{ZuHone.etal16}. It is important to note that these fields are \emph{not} turbulent since they are missing higher-order features like intermittency. The equivalent spectrum derivation makes the assumption of homogeneity (at \autoref{eqn:sf_variance_acf}) which could have significant consequence. However, this could also have small consequence as we are looking at second-order statistics only (autocorrelation, power spectrum, structure function) which are minimally influenced by the higher-order effects of intermittency and non-Gaussianity. The influence of intermittency on $\ESarg$ will be left for future work. Though, we do apply the technique to observations of the turbulent solar wind in \autoref{sec:examples}.


\subsubsection{Bias Validation}
We generate $D=1$ and $D=2$ fBm fields, sampling a range of $\alpha \in [D, D+2]$ corresponding to $\mathcal{E}_{D}(k)$ power law $\beta \equiv \alpha - D + 1 \in [1, 3]$ and $k_0 \in [10, 20]$. For $D=1$ we set $N=10^{6}$ and for $D=2$ we set $N=2048$ with $\mathcal{L} = 2\pi$.

\begin{figure}
    \centering
    \includegraphics[scale=1]{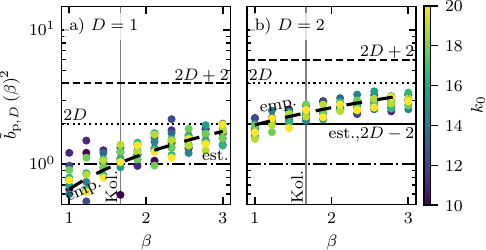}
    \caption{Similar to \autoref{fig:semi_analytical_peak_b}. The $b$ factor ($\bp$) corresponding to aligning the position of the peaks of $\ESarg$ and $\Earg$ using fBm fields with a spectrum of \autoref{eqn:exp_spectrum} for different $\beta$ and $k_0$ values (circles, with the colour bar representing different $k_0$). The black horizontal lines correspond to $1$ (dash-dot), $2D-2$ (solid), $2D$ (dotted), and $2D+2$ (dashed) as discussed in \autoref{sec:filter_function} with $\bpest$ denoted as ``est.'' where appropriate. The $\beta$-dependent thick black dashed line is the empirical formula \autoref{eqn:empirical_b_estimate} ($\bpemp$, denoted as ``emp.''). The solid vertical gray line represents the (Kolmogorov) power law slope of $-5/3$.}%
    \label{fig:numerical_b_estimate_1d_2d}%
\end{figure}

Performing a similar technique that was used to produce \autoref{fig:semi_analytical_peak_b}, we calculate $\Earg$ and $\ESarg$ for each fBm field and estimate the effective energy-containing scales $\kp$ and $\kep$. \autoref{fig:numerical_b_estimate_1d_2d} shows the estimate for $\bp$: the $b$-factor required to equate $\kep$ to $\kp$. We see that the fBm fields follow $\bpemp$. There is some noticeable scatter, likely due to the numerical differentiation and the binning and interpolation procedure.

\begin{figure}
    \centering
    \includegraphics[scale=1]{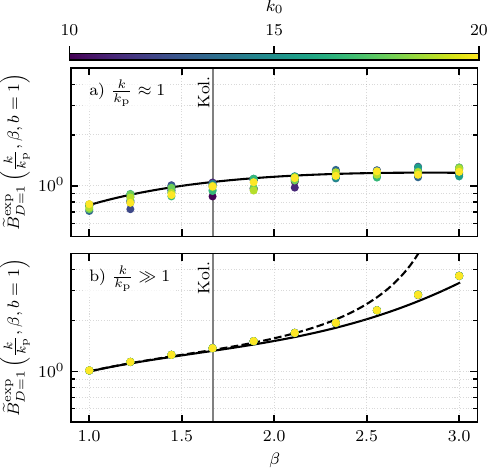}
    \caption{The ratio of the amplitude of $\ESarg$ to $\Earg$ of $D=1$ fBm fields of different $\beta$ and $k_0$ at (a) the peak $k,\ke \approx \kp$ and in (b) the power law regime $k,\ke \sim 30\kp$. The black solid line corresponds to the analytical $\Bexp$ (\autoref{sec:exp_analytical_bias}). The black dashed line corresponds to the analytical $\Bpow$ (\autoref{sec:pure_powerlaw_bias}). The solid vertical gray line represents the (Kolmogorov) power law slope of $-5/3$.}%
    \label{fig:numerical_B_estimate_1d}%
\end{figure}

\begin{figure}
    \centering
    \includegraphics[scale=1]{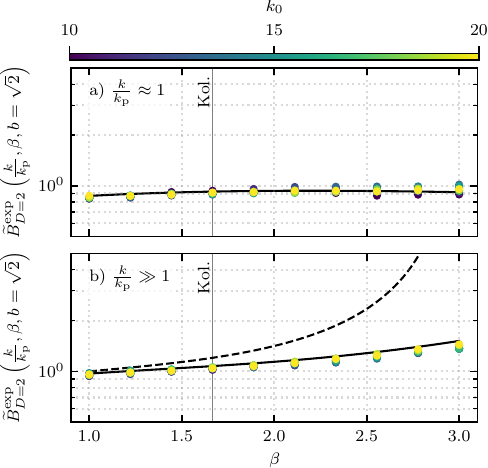}
    \caption{Similar to \autoref{fig:numerical_B_estimate_1d} for $D=2$. In panel (b) $k,\ke \sim 5 \kp$.}%
    \label{fig:numerical_B_estimate_2d}%
\end{figure}

\autoref{fig:numerical_B_estimate_1d} ($D=1$) and \autoref{fig:numerical_B_estimate_2d} ($D=2$) compare the amplitudes of $\ESarg$ to $\Earg$ of the fBm fields at the effective energy-containing scale (of $\Earg$: $k,\ke = \kp$) and in the asymptotic pure power law range ($k,\ke \gg \kp$). Values for the power law range are determined by selecting a single $k$ where $k,\ke \gg \kp$. The structure function is sampled with linear spacing $\Delta x$. The linear spacing of $\Delta x$ corresponds to $\ke$ with non-linear, non-integer spacing (proportional to $z = 1/n$ for $n \in [1, N/2]$) which leads to few data points in log-spaced bins at large $\ke$. This can result in poor numerical derivative estimates at the large $\ke$. By increasing the resolution (decreasing $\Delta x$) larger $\ke$ can be determined more accurately. Due to these resolution effects, numerical noise dominates at large $\ke$ so we are limited to $k,\ke = 30\kp$ for \autoref{fig:numerical_B_estimate_1d} and $k,\ke = 5 \kp$ for \autoref{fig:numerical_B_estimate_2d}. At larger $\ke$ than selected, without the numerical noise, we expect $\Bexp$ will approach $\Bpow$ (\autoref{sec:powerlaw_approx}). The fBm amplitude biases for both \autoref{fig:numerical_B_estimate_1d} and \autoref{fig:numerical_B_estimate_2d} match the analytical expression well with only a small amount of scatter in \autoref{fig:numerical_B_estimate_1d}(a).


Since the fBm fields match so well to the analytical expressions, we have confidence in an appropriate selection for $b$ and $\mathcal{B}$ to correct for the systematic biases in $\ESarg$. An initial estimate $\ESarg$ can be corrected for by choosing an analytical expression for $\mathcal{B}$. There is a bit of leeway in a suitable choice of $b$ as there is some scatter in \autoref{fig:numerical_b_estimate_1d_2d}. Either $b=\bpest$, or $b=\bpemp$ would be natural choices. We recommend $b=\bpest$ due to the simplicity. The factor $\bpemp$ requires prerequisite knowledge of $\beta$ or an additional correction step. At the Kolmogorov slope $\beta \sim 5/3$, $\bpest$ is an appropriate estimate to $\bpemp$.

\subsubsection{Large 1D fBm Validation}
As additional validation, we have created a large ($N = 10^{7}$) 1-dimensional fBm field using the spectral synthesis method (with $\mathcal{L}=2\pi$, $k_0=20$, $\beta=5/3$). We compare the bias (ratio) of $\ESarg$ to $\Earg$. The ``ground truth'', $\Earg$,  is a typical Fourier transform PSD (the Periodogram\cite{Schuster98}) calculated via the \texttt{FFT}. We test with two debiasing functions: $\Bpow$ and $\Bexp$.

\begin{figure}
    \centering
    \includegraphics[scale=1]{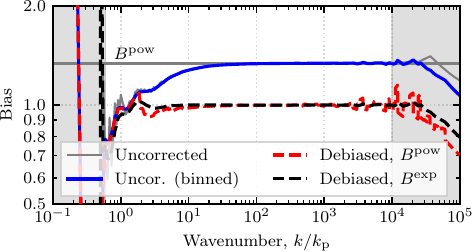}
    \caption{Bias ($\ESarg/\Earg$) of the equivalent spectrum $\ESarg$ compared to the ``ground truth'' \texttt{FFT}-based estimate for $\Earg$. The gray curve is the initial $\ESarg$ and the blue line shows the same $\ESarg$ but we have binned and interpolated to reduce the high-frequency noise. The gray horizontal line shows the constant pure power law bias $\Bpow(b=1, \beta=5/3)$. The red dashed line shows de-biasing using $\mathcal{B} = \Bpow(\betaest,b=1)$. The black dashed line corresponds to debiasing using $\mathcal{B}=\Bexp(k/\kp, b=1, \beta=5/3)$. The gray shaded regions indicate (roughly) where the errors become large.}
    \label{fig:large_validation}%
\end{figure}

\autoref{fig:large_validation} shows that the ``Uncorrected'' curve $\ESarg$ is noisy at $\ke \sim \kp$ and at $\ke \gtrsim 10^{3}$. This highlights the necessity of smoothing as the noise will propagate into $\betaest$ and then into $\mathcal{B}^{-1} \ESarg$. The binning we have applied is by-eye examined to ensure the important features (slopes and scales) are retained -- future work will focus on the case of intermittent signals as well as for signals with data gaps. The ``Uncorrected (binned)'' curve correctly reduces the noise at $\ke \sim \kp$. Our simple smoothing and binning steps are unable to significantly reduce the noise at $k \gtrsim 10^{4}$. The ``Uncorrected'' and ``Uncorrected (binned)'' spectrum follows $\Bpow(b=1,\beta=5/3)$ at $1 \lesssim k/\kp \lesssim 10^{4}$ as asserted by our analytical analysis in \autoref{sec:powerlaw_approx}.

The two ``Debiased'' $\mathcal{B}^{-1}\ESarg$ are functionally identical to $\Earg$ when $30 \lesssim k/\kp \lesssim 2\times10^{3}$. For all versions of $\ESarg$ (``Uncorrected'', ``Uncor. (binned)'', ``Debiased, $\Bpow$'', ``Debiased, $\Bexp$''), the bias at $\kp$ is small. As $\ke \rightarrow 0$ (which corresponds to $\ell \rightarrow \infty$) $\barS \rightarrow \mathrm{const.}$ which means $\d \barS/\d\ell \rightarrow 0$. We expect the $\ke \lesssim \kp$ region will not necessarily capture the exponential growth that we have modelled.
At $k/\kp \gtrsim 10^{3}$, the ``Debiased, $\Bpow$'' curve has large fluctuations around the expected value. This is due to the numerical noise for $\betaest$ propagating into $\Bpow$.

The results for ``Debiased, $\Bpow$'' appear equivalent to ``Debiased, $\Bexp$'', but ``Debiased, $\Bexp$'' theoretically only applies to the spectral model of the form \autoref{eqn:exp_spectrum}. For simplicity and robustness, in \autoref{sec:examples}, we will stick to debiasing using $\Bpow$ with a local power law estimate $\betaest$. The disadvantage of using $\betaest$ is that it is equivalent to using a second derivative of $\barS$ which results in the numerical difficulties at large $\ke$.

\section{Examples}
\label{sec:examples}

We apply $\ESarg$ to simulated and genuine data products with suitable resolution and for \red{dataset} dimensions $D=1,2,3$. For comparison, we generate a ``ground-truth'' $\Earg$ using a typical Fourier transform method (\ie{}Periodogram) via the \texttt{FFT}\cite{Schuster98, Stoica.Moses05}.

\red{\textbf{We now emphasize some additional important caveats.}} \red{The spectra, $\ESarg$ and $\Earg$, correspond to the spectra representing the \emph{dataset}. Usually, the dimension of the dataset $D$ is \emph{not} equal to the dimension of the system. In other words, observations typically obtain projections and slices of the full (usually 3D) system -- we have lost information. Nonetheless, the spectra, $\ESarg$ and $\Earg$, are associated with the \emph{dataset} and their associated biases, $b$ and $\mathcal{B}$, are assumed to be the same as investigated above with $D$ given by the dataset dimension.}

\red{The relations of the spectra of the dataset to the spectra of the system can be non-trivial. Additionally, aliasing issues can be introduced resulting from mis-interpreting Fourier modes in the observation that are associated with the un-observed dimensions. These issues are present in \emph{all} situations where the full system cannot be observed. We describe some of these cases for the examples that we show in this section in more detail in Appendix \ref{sec:sliced_and_projected_data}.}

Note that the analyses in this section are not intended to be a breakdown or discussion on PSD themselves, or provide any physical insights. We merely compare the equivalent spectrum to traditionally accepted techniques on high-fidelity data across a range of data-types.

\subsection{Solar Wind, $D=1$}
\label{sec:solar_wind_example}


Viewed as a natural wind tunnel, the solar wind as a ``laboratory'' for plasma turbulence experiments \cite{Marsch91, Bruno.Carbone13, Verscharen.etal19} is unique as it provides decades of high-resolution in-situ astrophysical plasma turbulence observations. As a result, the solar wind spectrum is remarkably well studied \cite{Kiyani.etal15}. The solar wind has also been used in previous analyses of the equivalent spectrum \cite{Chhiber.etal18, Thepthong.etal23}.

\red{Solar wind obervations are performed by sensors that generate a time-series of data. In other words, the observations have a dataset dimension of $D=1$ (all time-series observations have dataset dimension $D=1$). Of course, the solar wind magnetic field fluctuations are determined by a system that is spatially \emph{not} 1D \ie{}the system dimension $>1$. Given some additional assumptions (such as Taylor's hypothesis), the time-series observations can be assumed to approximate a 1-dimensional spatial slice of the physical system. We discuss additional intricacies with estimating spectra associated with $D=1$ observed slices in Appendix \ref{sec:sliced_and_projected_data}.}

\red{Some solar wind observations rely on multiple spacecraft such as the Magnetospheric Multiscale (MMS) mission \cite{Burch.etal16}. The unique measurement capabilities of MMS, which are a set of satellite that fly in formation, enables direct access to measurements with a handful of fixed spatial separations. These spatial separations probe specific equivalent wavenumbers of the spectrum using the same data (which would not be possible to estimate using traditional Fourier techniques). This application is an advantage of the \emph{equivalent spectrum} \cite{Chhiber.etal18}.}


We use data from \emph{Wind}\cite{WilsonIII.etal21} which has been at Lagrange point 1 ($\approx 1\,\mathrm{au}$) since May 2004. One month [2007/01/01--2007/02/01] of calibrated version 4 (datatype ``h4-rtn'') solar wind magnetic field data from the \emph{Wind} magnetic field instrument (MFI) is obtained using \href{https://github.com/spedas/pyspedas}{PySPEDAS}. The data are $D=1$ time series observation that has $N=28,955,151$ data points for each magnetic field component $\vec{b}(t) = (b_{\mathrm{R}}(t), b_{\mathrm{T}}(t), b_{\mathrm{N}}(t))$ in units of $\mathrm{nT}$, with a sampling spacing of $\Delta t \approx 0.092\,\mathrm{s}$. We obtain the fluctuations by $\delta \vec{b}(t) = \vec{b}(t) - \avg{\vec{b}(t)}$ where $\avg{\cdot}$ is the temporal average over the entire interval. We apply a $1\%$ Tukey window to each $\delta \vec{b}(t)$ to reduce aliasing effects \cite{Matthaeus.Goldstein82}. The structure function, its corresponding PSD estimates and the Periodogram estimates are calculated on this windowed data.


\begin{figure*}
    \centering
    \includegraphics[scale=1]{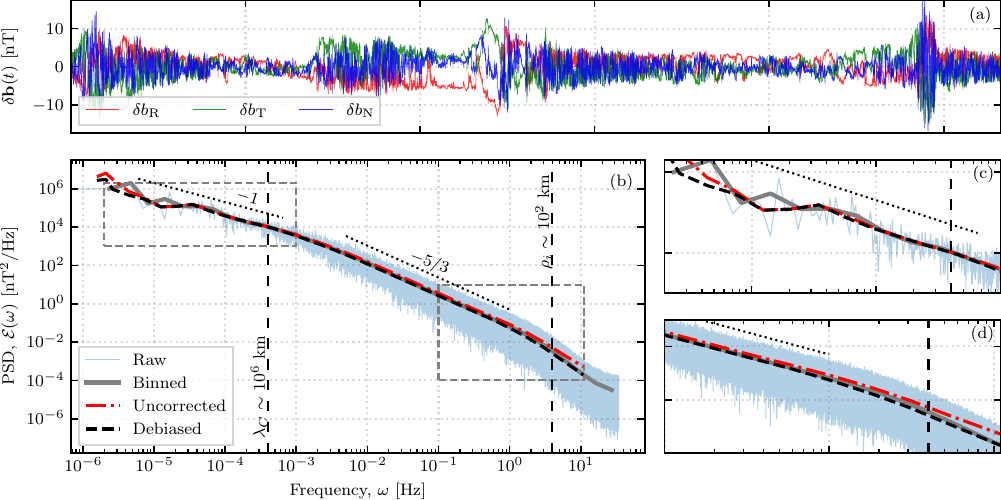}
    \caption{(a) 1 month of \emph{Wind} solar wind magnetic field component fluctuation data. (b) PSD estimate (Periodogram) of the data in (a) raw (solid blue), and binned (solid gray). The dash-dot red line indicates the equivalent spectrum $\ESarg$ with $b=1$. The black dashed line shows $\ESarg$, we have debiased the initial estimate of $\ESarg$ by using $\mathcal{B} = \Bpow$ and $\betaest$. The black dashed vertical lines show the approximate scales where the spectrum is expected to change slope: the correlation scale $\lambda_C$, and the ion gyro-radius $\rho_i$. (c), (d) show zoomed-in insets from (b) as indicated by the gray dotted boxes near to $\lambda_C$ and $\rho_i$ respectively.}%
    \label{fig:SW_example}%
\end{figure*}


The solar wind data (shown un-windowed in \autoref{fig:SW_example}(a)) has Fourier PSD estimates according to solar wind turbulence phenomenology at $1\,\mathrm{au}$ which are shown in \autoref{fig:SW_example}(b,c,d) with the following distinct regions of interest:
\begin{itemize}
    \item At the largest scales (smallest frequency, $\omega$) the solar wind PSD measures the temporal variability of the source of the solar wind (the Sun) and exhibits a $\omega^{-1}$ power law \cite{Wang.etal24a, Dorseth.etal24}. This is often called the $1/f$ or pink noise range. In this region, since $\beta \approx 1$, $\Bpow$ is minimal (see \autoref{sec:pure_powerlaw_bias}). Therefore, the ``Uncorrected'' and ``Debiased'' estimates are similar, if not identical. See \autoref{fig:SW_example}(c) for an inset/close-up. The ``Uncorrected'' and ``Debiased'' estimates closely follow the ``Binned'' $\Earg$ because $\Bpow$ is $\approx 1$.

    The correlation length, $\lambda_{\mathrm{C}}$, is an outer/energy-containing scale \cite{Pope00} that represents the largest typical scale over which the turbulent fluctuations are correlated. The correlation length $\lambda_{\mathrm{C}}$ varies $\sim \lambda_{\mathrm{C}}^{-1/2}$ as the radial distance increases from the sun \cite{Bishop.etal25} and also over sample and time \cite{Wrench.etal24}, but typically, is of the order $\sim 10^{6}\,\mathrm{km}$ at $1\,\mathrm{au}$ \citep{Matthaeus.etal05, Cuesta.etal22}. The spectrum follows $\omega^{-1}$ scaling for $\omega < U/\lambda_{\mathrm{C}} \sim 0.0004\,\mathrm{Hz}$.

    
    At scales $\omega \gtrsim U/\lambda_{C}$, the $1/f$ range transitions to a turbulence inertial range.

    \item The turbulence inertial range is where the classic MHD turbulence phenomenology is observed with a $\omega^{-5/3}$ power law corresponding to the Kolmogorov turbulence phenomenology \cite{Kolmogorov41}. The solar wind is a highly dynamic environment so depending on the realization associated with the observation dataset, departures from Kolmogorov $\beta = 5/3$ can be observed \cite{Tessein.etal09, Boldyrev.etal11, Horbury.etal08, Chen16}. The spectrum begins to steepen past $\omega \gtrsim U/\lambda_{\mathrm{C}}$ and transitions to a $\beta \sim 5/3$ cascade region. When the spectrum is steeper there is a small bias in amplitude of the ``Uncorrected'' estimate. Yet, the behaviour of the ``Uncorrected'' curve remains consistent with $\Earg$ \ie{}the correct power law is obtained.
    
    At first glance, it may appear that the ``Uncorrected'' power law slope is shallower than $5/3$. As discussed in \autoref{sec:exp_analytical_bias} and \autoref{sec:powerlaw_approx}, $ \ESarg$ will have the correct power law at $\omega \gtrsim 10 \omega_{\mathrm{p}}$ where $\omega_{\mathrm{p}}$ is the effective energy-containing scale (in this case, $\sim U/\lambda_{\mathrm{C}}$). In the inertial range, at large $\omega$, $\betaest \sim 5/3$.
    
    The ``Debiased'' spectrum almost exactly follows the ``Binned'' $\Earg$. The bias in the ``Uncorrected'' curve is well described by $\Bpow$ using $\betaest$.

    

    The inertial range is bounded at the largest scale by $\lambda_{\mathrm{C}}$ and the smallest scale by dissipation effects. In the nearly-collisionless solar wind, a relevant dissipation scale is the ion-gyro scale $\rho_i$.
    
    \item In the solar wind, the kinetic effects modify the cascade at scales comparable to the ion kinetic scale. When the plasma beta $\geq 1$ this scale is the ion-gyroradius. At $1\,\mathrm{au}$, at the scales where the ions (protons) gyrate around the magnetic field are $\rho_i \sim 10^{2}\,\mathrm{km}$ \cite{Chen.etal14, Franci.etal16, Wang.etal18}, the MHD fluid description fails and kinetic effects begin to dominate. From Taylor's hypothesis $U/\rho_i \sim 4\,\mathrm{Hz}$. The dynamics in this region are debated and depend on a wide range of conditions, but overall, a steepening of the spectrum (from $5/3$) is expected \cite{Marsch06, Kiyani.etal15}.

    \autoref{fig:SW_example}(b,d) shows that the spectrum departs from $\beta \sim 5/3$ as $\omega \gtrsim U/\rho_i$. The amplitude bias in the ``Uncorrected'' curve becomes larger. Once again, this effect is due to the intrinsic bias in the amplitudes of the spectral estimate for power law behaviour that is corrected for when debiased. The ``Debiased'' estimate yields a similar result to the binned periodogram estimate.

    At larger $\beta$, and at the transition region around $\omega \sim U/\rho_i$, using $\Bpow$ with $\betaest$ still adequately describes the bias in the ``Uncorrected'' curve.
\end{itemize}

For the regions where the bias is $\approx 1$ (\ie{} where the spectrum is $\sim \omega^{-1}$), the ``Uncorrected'' estimate well represents the $\Earg$. Since the behaviour of the spectrum is almost entirely power law, the choice of $b$ is not significant for the outcome of the ``Debiased'' curve but for a consistent argument we have chosen $b=1$ ($\best$) and applied the appropriate debiasing function $\Bpow$. Overall, the ``Uncorrected'' curve yields similar behaviour to $\Earg$ with a small amplitude error in the $\omega^{-1}$ and $\omega^{-5/3}$ power law regimes respectively. The ``Debiased'' $\mathcal{B}^{-1}\ESarg$ closely follows $\Earg$ at all scales with minimal error. Our spectral estimates are consistent with literature and expected solar wind turbulence phenomenology \cite{Kiyani.etal15}.

\subsection{Interstellar Medium, $D=2$}
\label{sec:ISM_example}

Telescope observations using wavelengths from radio-waves to gamma-rays often provide 2-dimensional images of gas, dust, and plasma. Spectral analyses on these images have been used to understand the dynamics of \eg{}the Milky Way's interstellar medium (ISM) \cite{Elmegreen.Scalo04, Koch.etal19, Burkhart21}, the ISM of other galaxies \cite{Li.etal20, Ganguly.etal23, Gerrard.etal23}, the intracluster medium of galaxy clusters \cite{Churazov.etal12, Simionescu.etal19, Zhuravleva.etal19, Romero.etal23}, and the cosmic microwave background \cite{Efstathiou04, Sullivan.etal25}.

The Local Group is a prime candidate for understanding processes in galaxies \cite{Livio.Brown06}, enabling exceptionally high-fidelity data from our neighbourhood (astronomically speaking). Within our Local Group, the Large Magellanic Cloud (LMC) is an irregular dwarf galaxy with a single spiral arm \cite{Ruiz-Lara.etal20} that spans an angular diameter distance of $\approx 10.7^\circ$ in the sky \cite{Cook.etal14}. We use already processed available data which makes use of a spectral feathering technique combining data from a range of telescopes to provide a higher-resolution observation and therefore spectral range \cite{Clark.etal21, Clark.etal22}. We use the Herschel Photodetector Array Camera and Spectrometer (PACS) \cite{Poglitsch.etal10} $160\,\mu \mathrm{m}$ band \cite{Clark.etal22}. Observations of far-infrared radiation corresponds to thermal emission from dust heated by starlight \cite{Werner.etal78} and can be used to constrain dust properties such as surface density and temperature \cite{Gordon.etal14, Clark.etal23}.





\begin{figure*}
    \centering
    \includegraphics[scale=1]{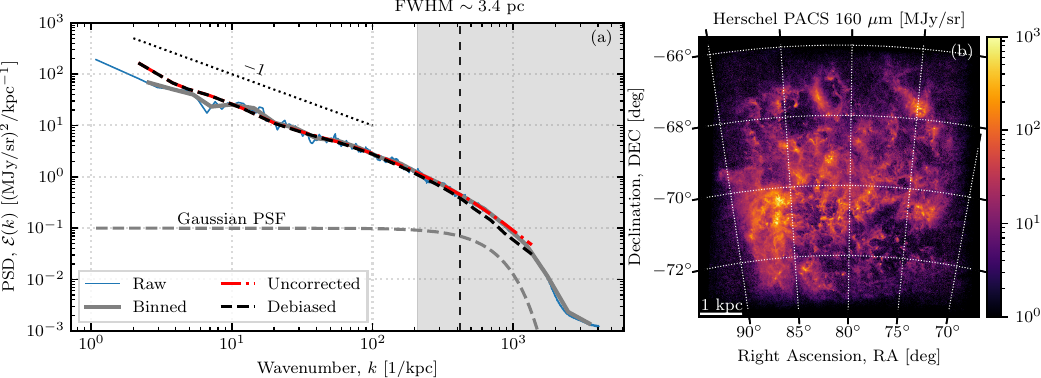}
    \caption{Similar to \autoref{fig:SW_example}; (a) spatial PSD estimates of (b) Herschel PACS $160\,\mu\mathrm{m}$ observation of the Large Magellanic Cloud \cite{Clark.etal21, Clark.etal22}. The gray shaded region indicates where the effects of the limiting resolution of the data (the telescope's point spread function) starts to influence the spectral estimates ($k \sim \frac{\sqrt{2}}{2\times\text{FWHM}}$). Also showcasing the PSF influence, the gray dashed line shows a Gaussian approximation for the PSF normalized to an order of magnitude below unity. We have used $b=\sqrt{2}$.}%
    \label{fig:LMC_example}%
\end{figure*}

Data with sharp boundaries can introduce aliasing artefacts into the Fourier transform \cite{Brault.White71}. Astronomical literature typically reduces these artefacts via window functions \cite{Matthaeus.Goldstein82, Dickey.etal01, Khatri.Gaspari16} or by estimating the PSD by the $\Delta$-variance \cite{Stutzki.etal98, Bensch.etal01, Ossenkopf.etal08, Ossenkopf.etal08} or difference-of-Gaussian \cite{Arevalo.etal12, Zhuravleva.etal14a, Zhuravleva.etal19} techniques. We take a central $9000^2$ section of the provided foreground subtracted data and apply a $25\%$ Tukey window to reduce sharp boundaries. Seen in \autoref{fig:LMC_example}(b), the image is effectively padded with 0's and smoothed at the edges of the LMC.



\autoref{fig:LMC_example}(a) shows the PSD estimates for the data \red{$s_{2}^{\mathrm{obs}}(\vec{x})$ (with dataset dimension $D=2$ and $\vec{x} \in \mathbb{R}^2$)} in \hyperref[fig:LMC_example]{(b)}. Literature spectral analysis of the LMC over a range of wavelengths and telescopes provides a consistent power law of $\beta \sim 1$ \cite{Koch.etal20}. This is compatible with our analysis. ISM analyses typically use the angle-\emph{averaged} spectrum \red{which is obtained by averaging over a spherical shell $\mathcal{S}(k)$ of radius $k = \abs{\vec{k}}$,
\begin{align}
    \label{eqn:angle_averaged_spectrum}
    \overline{E}_{2}(k) = \frac{1}{2 \pi k} \int E_{D}(\vec{k}) \d \mathcal{S}(k),
\end{align}
and is related to the angle-integrated spectrum by $\overline{E}_{2}(k) \propto k^{-1} \mathcal{E}_{2}(k)$ (see \autoref{eqn:integrated_spectrum_from_modal})}. Therefore, reported power laws in literature are steeper by $k^{-1}\,$ \cite{Elmegreen.etal01, Elmegreen.etal03, Koch.etal20}.


The gray region in \autoref{fig:LMC_example}(a) indicates the region where the point spread function (PSF) starts to dominate the spectral estimate. For the LMC, it has been acknowledged that apparent spectral breaks are caused by the PSF \cite{Koch.etal20}. The observed field $s_{2}^{\mathrm{obs}}(\vec{x})$ is a convolution of the non-blurred field $s_{2}^{\mathrm{true}}(\vec{x})$ with the PSF of the telescope $G^{\mathrm{PSF}}_{2}(\vec{x})$ (we are ignoring the influence of noise),
\begin{align}
    s_{2}^{\mathrm{obs}}(\vec{x}) = \bracket{s_{2}^{\mathrm{true}} \ast G^{\mathrm{PSF}}_{2}}(\vec{x}).
\end{align}

\red{Note that $s_{2}^{\mathrm{true}}(\vec{x})$ is not necessarily the field that represents the system. \emph{If} the system is 3D, then $s_{2}^{\mathrm{true}}(\vec{x})$ is the corresponding projection (integration) along the line-of-sight and the observation, $s_{2}^{\mathrm{obs}}(\vec{x})$, applies a convolution with the telescope PSF on top of the projection of the field. The spectral estimations we show represent the dataset. Conversion to the 3D system is non-trivial and requires further assumptions that are out-of-scope for this paper (see \autoref{sec:sliced_and_projected_data} for more details).}

The convolution theorem states that a convolution in real-space is multiplication in Fourier space,
\begin{align}
    \label{eqn:obs_psf_psd}
    E^{\mathrm{obs}}_{2}(\vec{k}) = E^{\mathrm{true}}_{2}(\vec{k}) E^{\mathrm{PSF}}_{2}(\vec{k}).
\end{align}
where $E^{\mathrm{true}}_{2}(\vec{k})$ is given by \autoref{eqn:fourier_spectrum_defn}. If we assume isotropy of all the spectra in \autoref{eqn:obs_psf_psd}, by integrating both sides of \autoref{eqn:obs_psf_psd} over regions where $k = \abs{\vec{k}}$,
\begin{align}
    \mathcal{E}_{2}^{\mathrm{obs}}(k) = \mathcal{E}^{\mathrm{true}}_{2}(k) \overline{E}^{\mathrm{PSF}}_{2}(k),
\end{align}
where $E^{\mathrm{PSF}}_{2}(k)$ is the angle-averaged spectrum (\autoref{eqn:angle_averaged_spectrum}) of the point-spread function and $\mathcal{E}_{2}^{\mathrm{obs}}(k)$, $\mathcal{E}^{\mathrm{true}}_{2}(k)$ are the observed and non-blurred angle-integrated spectra (\autoref{eqn:integrated_spectrum_from_modal}) respectively. Therefore, our observed PSD is a multiplication of the non-blurred angle-integrated spectra and the angle-averaged spectrum of the PSF. Modelling the PSF of the Herschel PACS $160\,\mu \mathrm{m}$ band as a Gaussian with a full-width at half-maximum of $\approx 3.4\,\mathrm{pc}$ \cite{Aniano.etal11}, we show $\overline{E}^{\mathrm{PSF}}_{2}(k) = e^{-k^2 \sigma^2}$ in \hyperref[fig:LMC_example]{(a)} scaled to an order of magnitude below the amplitude for comparison purposes only. We do not retrieve $\mathcal{E}_{D}^{\mathrm{true}}(k)$ in our analysis as the Gaussian PSF we use is not the true PSF present in the data \cite{Aniano.etal11, Clark.etal21} and the process for obtaining $\mathcal{E}_{D}^{\mathrm{true}}(k)$ is the same for $\ESarg$ as traditional $\Earg$. See other spectral analysis literature for removing the influence of the PSF \cite{Churazov.etal12, Koch.etal20}.

The ``Uncorrected'' slope at $k \gtrsim \sqrt{2}/\bracket{2\times\text{FWHM}}$ follows the ``Binned'' periodogram. The spectrum is steeper in this range resulting in a larger bias and therefore a ``Debiased'' slope that is smaller in amplitude at $k \gtrsim \sqrt{2}/\text{FWHM}$ than the ``Binned'' spectrum. The exponential-like roll-off in the (approximately Gaussian) PSF means the local power law de-biasing technique might not be as effective. In this region, numerical noise for $\betaest$ could also be strong, similar to \autoref{fig:large_validation}. This could explain the discrepancy between the ``Debiased'' and ``Binned'' curves. 

The ``Uncorrected'' and ``Debiased'' PSD estimates are similar where $k \lesssim \frac{\sqrt{2}}{2*\text{FWHM}}$ due to the $\sim k^{-1}$ power law behaviour which has small bias. The ``Uncorrected'', and ``Debiased'' slopes are similar to the ``Binned'' periodogram estimate beyond the FWHM scale which means the same techniques for reducing the influence of the PSF could be applied to $\ESarg$ as $\Earg$. The PSF effects on the spectrum are outside the scope of this paper.


\subsection{Isotropic Hydrodynamic Simulations, $D=3$}


In the absence of any real observation of turbulence that spans an entire 3-dimensional domain, we resort to simulations. We use data from an available high-resolution ($1024^3$) 3-dimensional incompressible, homogeneous, isotropic hydrodynamic turbulent simulation \cite{Cardesa.etal17}. The simulation is a direct numerical simulation of turbulence in a triply periodic cube where the Navier-Stokes equations are solved for in an incompressible fluid using a deterministically forced and statistically steady pseudo-spectral code \cite{Orszag.Patterson72}. We use a single snapshot -- \texttt{hit.1024.06468.h5} -- which corresponds to $\approx 13$ integral time-scales. The turbulence is fully developed.

\begin{figure*}
    \centering
    \includegraphics[scale=1]{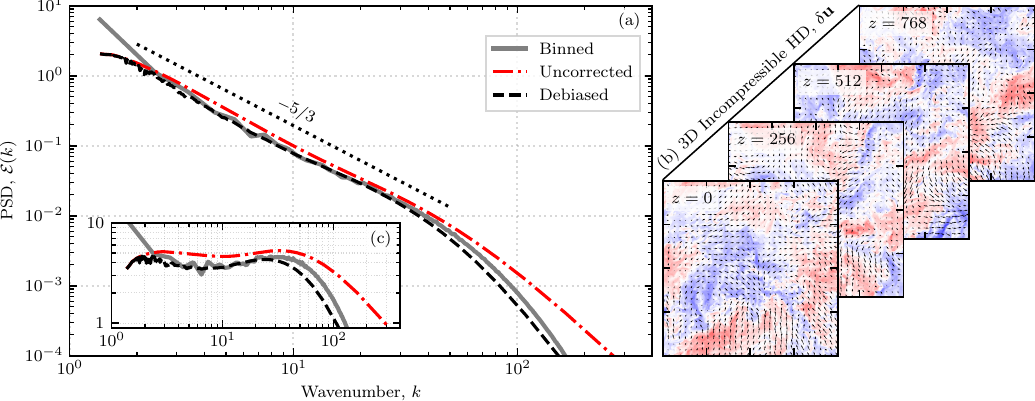}
    \caption{Similar to \autoref{fig:SW_example}; (a) spatial PSD estimates of a $1024^3$ 3D incompressible hydrodynamic turbulence simulation\cite{Cardesa.etal17} using $b=2$. (b) $z$-slices of the simulation cube with the colour map representing $\delta u_{z}$ from -7 (blue) to 7 (red) and the vector field representing $\delta u_{x}, \delta u_{y}$. (c) the same spectra shown in (a) but compensated by $k^{5/3}$. Axes scales are in code units.}%
    \label{fig:3D_incompressible_1024_HD}%
\end{figure*}

To ease computations, we calculate the following parallel, and perpendicular structure functions using the simulation's velocity field fluctuations $\delta \vec{u}(\vec{x}) = (\delta u_x(\vec{x}), \delta u_y(\vec{x}), \delta u_z(\vec{x}))$ (depicted in \autoref{fig:3D_incompressible_1024_HD}(b)),
\begin{subequations}
    \begin{align}
        S^{\parallel,x}_{3}(\ell) &= \avg{\abs{\delta u_{x}(\vec{x}) - \delta u_{x}(\vec{x} + \ell \unitvec{e}_{x})}^2},\\
        \label{eqn:isotropy_y}
        S^{\perp,x}_{3}(\ell) &= \avg{\abs{\delta u_{y}(\vec{x}) - \delta u_{y}(\vec{x} + \ell \unitvec{e}_{x})}^2},\\
        \label{eqn:isotropy_z}
        &\approx \avg{\abs{\delta u_{z}(\vec{x}) - \delta u_{z}(\vec{x} + \ell \unitvec{e}_{x})}^2},
    \end{align}
\end{subequations}
to form $\overline{S}_{3}(\ell) = S^{\parallel,x}_{3}(\ell) + 2 S^{\perp,x}_{3}(\ell)$. Note, \autoref{eqn:isotropy_y} and \autoref{eqn:isotropy_z} are equal only under perfect isotropy. 

Since the simulation is incompressible, the zero-divergence condition required for this decomposition is met. The ``Binned'' $\Earg$ (ground-truth) is calculated by integrating partial shells in the quadrant of positive $k_x, k_y$, with all of $k_{z}$. The resultant $\Earg$ is multiplied by $4$ to account for the missed quadrants for $k_{x}, k_{y}$. The PSD estimates using $b=2$ ($\bpest$) are shown in \autoref{fig:3D_incompressible_1024_HD}(a).

A Kolmogorov $k^{-5/3}$ cascade region is present in the simulation. We see the $k^{-5/3}$ inertial range (power law slope) is captured well by $\ESarg$ (for both ``Uncorrected'' and ``Debiased''). The ``Uncorrected'' curve has a larger amplitude in the inertial range ($3 \lesssim k \lesssim 20$) than $\Earg$. This amplitude is appropriately corrected for in the ``Debiased'' curve as it almost exactly follows the ``Binned'' estimate. Due to the nature of the forcing in the simulation, there is no observed region with exponential growth (like that which was modelled in \autoref{sec:exp_analytical_bias}).

At larger $k$, the viscous term in the Navier-Stokes equation begins to dominate and results in an exponential roll-off in the spectrum \cite{Foias.etal90, Terry.etal12}. Towards the dissipation regime $k \gtrsim 60$, the ``Uncorrected'' estimate has a large bias and shows steepening of the curve, but not enough to describe the expected decay of the ``Binned'' $\Earg$. At these large $k$, with steep decay curves, we would expect $\ESarg \sim k^{-3}$ corresponding to the steepest $\beta$ possible for the equivalent spectrum (\autoref{sec:pure_powerlaw_bias}). This is also seen in the structure function \cite{Monin.Iaglom75, Stutzki.etal98, Lazarian.Pogosyan06, Lazarian.Pogosyan08}. In a steep dissipation regime like $\Earg \sim e^{-k/k_\mathrm{diss}}$, $\betaest$ will approach 3 for $k \gtrsim k_{\mathrm{diss}}$. Since $\lim_{\beta\rightarrow 3^{-}}\Bpow(\beta, b) = \infty$, using $\mathcal{B} = \Bpow(\betaest, b)$ will result in decay of $\mathcal{B}^{-1}\ESarg$ that is more rapid than the expected $e^{-k/k_{\mathrm{diss}}}$. The accuracy of the ``Debiased'' $\mathcal{B}^{-1}\ESarg$ for steep dissipation regimes is not guaranteed. We see the ``Debiased'' $\mathcal{B}^{-1}\ESarg$ decays more rapidly than $\Earg$ at $k \gtrsim 40$. At $k \gtrsim 100$, the ``Uncorrected'' curve has approached $\betaest \approx 3$ and due to the clamping of $\betaest$ (\autoref{sec:implementation}), the ``Debiased'' curve will exhibit a constant power law behaviour.

Due to limited available bandwidth in the simulation, the $k^{-5/3}$ cascade region can be small \cite{Wan.etal10}. One such reason is the ``bottleneck effect'': energy can pile-up in $k$-space around the dissipation scale $k_{\mathrm{diss}}$ \cite{Falkovich94}. This is evident in \autoref{fig:3D_incompressible_1024_HD}(c) showing the compensated spectra $k^{5/3} \Earg$, the pile-up starts around $k \gtrsim 20$. It has been shown that the scaling properties of $\barS$ and $\Earg$ can be quite different in the bottleneck region \citep{Lohse.Muller-Groeling95, Lohse.Muller-Groeling96, Donzis.Sreenivasan10}. It has also been suggested that the second-order structure function is less sensitive to the bottleneck effect due to the non-local influence of the filter on $\Earg$ in \autoref{eqn:spectral_representation} \cite{Dobler.etal03}. The bottleneck effect in the ``Debiased'' curve appears less pronounced in \autoref{fig:3D_incompressible_1024_HD}(a). However, in \hyperref[fig:3D_incompressible_1024_HD]{(c)} the shape of compensated spectrum for the ``Debiased'' curve is almost identical for $k > 3$ to the ``Binned'' $k^{5/3}\Earg$ with just an apparent offset in equivalent wavenumber by $\approx 25\%$. Understanding the dynamics of the compensated equivalent spectrum (and other transformations on $\ESarg$) will be left for future work.

Overall, the ``Uncorrected'' curve shows that the equivalent spectrum must be debiased appropriately. In the case of a 3D incompressible hydrodynamic isotropic homogeneous simulation, using $\mathcal{B} = \Bpow$ with a local power law slope estimate $\betaest$ results in an appropriate estimation of the PSD. The inertial range is shown in both the ``Uncorrected'' and ``Debiased'' curves as $\Bpow$ is small for $\beta \sim 5/3$ and $D=3$. It is important to debias the equivalent spectrum since the ``Uncorrected'' curve looks like a broken power law spectrum of $\sim k^{-5/3}$ transitioning to $\sim k^{-3}$ at $k \gtrsim 60$. The ``Debiased'' curve is able to show dissipation-like behaviour beyond $k \gtrsim 60$. Although, it is unable to exactly match $\Earg$.

\subsection{\red{Missing Data}}

\begin{figure*}
    \centering
    \includegraphics[scale=1]{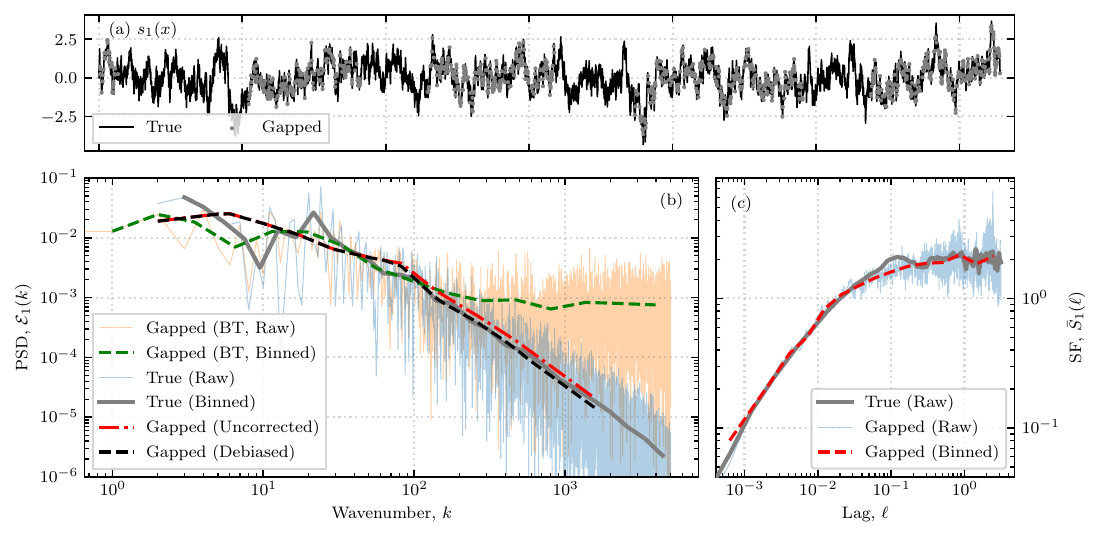}
    \caption{\red{(a) complete realization of a synthetic $D=1$ fBm field (black) and a mock sparse/gapped sampling  corresponding to $\approx 90\%$ missing data (gray dots). (b) PSD estimates on ``Gapped'' compared to the PSD estimates computed on the ``True'' fBm field (blue, and gray for the corresponding binned). (c) second order structure functions on the full data (gray) and on the gapped data (blue). The SF of the gapped data is binned to produce the smoothed SF estimate (red).}}%
    \label{fig:fbm_1d_gapped}%
\end{figure*}

\red{One of the main strengths of this method is its application to datasets with missing data, which Fourier methods are well-known to struggle with. We plan to test this more exhaustively in future work, however here we show an initial test as proof of concept.}

\red{We test our method on sparse/gapped synthetic $D=1$ fractional Brownian field (fBm) data. We synthesize an fBm field of $N=10000$ points in arbitrary units using the method described in \autoref{sec:fbm_validation} -- this makes the ``True'' data which we compare the estimates to. We generate ``Gapped'' data by randomly removing large blocks, as well as randomly removing singular points of data from the ``True''. We remove a total number of data points corresponding to $\approx 90\%$ missing data and $N=1000$ points (shown in \autoref{fig:fbm_1d_gapped}(a)). From the ``Gapped'' and ``True'' dataset, we can compute the ``True'' PSD estimate, and estimates of the PSD using only the ``Gapped'' data.}

\red{In \autoref{fig:fbm_1d_gapped}(c), we compute the SF of the gapped data and compare it to the SF of ``True''. The SF can be computed directly on the ``Gapped'' data where lag separations are available \ie{}values are based off pairs in the data, so if one or both values in a pair are missing, it can simply be ignored. It is recommended to apply corrections to ``Gapped'' SF estimations, however for our example we assume these corrections should be small since the field is Gaussian \cite{Wrench.Parashar24}. At small $\ell$, the SF shows the same power-law for both the ``Gapped'' and ``True''. The ``Gapped (Raw)'' estimate is noisy at large scales which neccessitates the binning procedure which we show as ``Gapped (Binned)''. The ``Gapped (Binned)'' SF mostly follows the ``True'' SF except at scales $0.05 \lesssim \ell \lesssim 0.2$ -- associated with the correlation scale of the synthesized fBm field. We use the ``Gapped (Binned)'' SF for computing $\ESarg$.}

\red{In \autoref{fig:fbm_1d_gapped}(b), we show the ``Uncorrected'' $\ESarg$ and ``Debiased'' $\mathcal{B}^{-1}\ESarg$ which corresponds to a PSD estimation solely from the ($\approx 90\%$) ``Gapped'' dataset. This constitutes a reasonable approximation of the ``True (Binned)'' (Periodogram) Fourier estimate on the full dataset. The `Debiased'' estimate follows the amplitude and slope in the power-law regime ($k \gtrsim 10^{2}$) as well as following the general trend of the ``True (Binned)'' PSD at $k \lesssim 10^{2}$.}

\red{For comparison of gapped PSD estimates, we include the Blackman-Tukey (BT) method \citep{Blackman.etal60}. The BT method estimates the PSD using the Fourier transform of an estimate for the autocorrelation function, hence, for the same reasons we can compute the structure function on the ``Gapped'' dataset, we can compute an estimate for the autocorrelation function on the ``Gapped'' dataset \ie{}only pairs of non-missing datapoints are considered. The BT method only exhibits similar behaviour to the ``True'' estimate at $k \lesssim 10^{2}$, the information at large $k$ ($k \gtrsim 10^{2}$) is lost. The BT performs poorly as a PSD estimate in this range whereas the equivalent spectrum method is able to obtain the expected behaviour.}

\red{However, further work is required for quantifying the accuracy of $\ESarg$ on gapped data of different types and where the data is non-Gaussian. Additionally, corrections to the estimation of the SF with gapped data should be considered in future work \cite{Wrench.Parashar24}.}

\section{Conclusion}

We have introduced a method for estimating the angle-integrated power spectral density $\Earg$ using the angle-averaged second-order structure function $\barS$: $\ESarg$ (\autoref{eqn:equivalent_spectrum}). We have thoroughly validated $\ESarg$ as a PSD estimate through a range of analytical and numerical techniques. With generalizations for dimension and power law of the data, we make the technique applicable to a wide range of fields and turbulence regimes.

Through $\ESarg$, we are able to investigate the relationship between Fourier wavenumber $k$ and lag/physical-scale $\ell$ through the parameter $b = k\ell$. Appropriate decisions on $b$ rely on the structure of $\Earg$, but we have provided estimated and empirical formulae that should be appropriate for many circumstances (\autoref{eqn:bp_est}, \autoref{eqn:empirical_b_estimate}). These results shed some light on the range of $b$ values employed in the literature \cite{Davidson.Pearson05, Davidson15, Colbrook.etal17, Squire.Hopkins17, Federrath.etal21, Thepthong.etal23}.

The ratio in amplitude between the structure function-based estimate for the spectrum $\ESarg$ and the actual spectrum $\Earg$, denoted $\mathcal{B}$, has also been investigated. We show that in most cases we are able to account for the differences. We have suggested techniques for reducing these effects. By modelling spectra analytically, or through fractional Brownian motion fields the appropriate $\mathcal{B}$-factor can be understood. We also propose a general purpose debiasing method by using the pure power law bias $\Bpow$ (\autoref{eqn:powerlaw_bias}) with a local power law estimate $\betaest$ (\autoref{eqn:local_powerlaw_est}) which works well on practical data, or where the expected power law is not well known. In some cases $\mathcal{B} \approx 1$ and this debiasing makes little difference; in other cases it makes a significant difference.

This paper relates previous literature usage of the power-law approximation of the equivalent spectrum \autoref{eqn:powerlaw_ESF_approx}\cite{Chasapis.etal17, Chhiber.etal18, Parashar.etal18, Wrench.Parashar24} to \autoref{eqn:equivalent_spectrum}\cite{Thepthong.etal23} in \autoref{sec:powerlaw_approx}. We also develop an understanding of the systematic biases of the equivalent spectrum through $\mathcal{B}$ which is previously unreported.

The behaviour of $b$ and $\mathcal{B}$ depending on the structure of $\Earg$ has been discussed throughout \autoref{sec:filter_function}.

Finally, we showcased use of the equivalent spectrum for 1-dimensional in-situ solar wind turbulence measurements, 2-dimensional dust surface density of the Large Magellanic Cloud, and a 3-dimensional hydrodynamic turbulence simulation. We find good agreements with Fourier PSD, which is used as the ground truth.

\red{We show an example of the equivalent spectrum performing with gapped and sparse data, but future work will require further investigation and quantification of the accuracy of the estimator.} Another area of potential improvement is in improving the technique by reducing noise introduced from the numerical differentiation and our naive binning and interpolation routines. Improvements could be made for example, through \eg{}Kalman \cite{Sarkka13, Kim.etal18}, Savitzky-Golay \cite{Persson.Strang03} or other filtering techniques \cite{VanBreugel.etal20, Schmid.etal22} and even semi-parametric profile fitting \cite{Wang.etal24}.


\appendix
\section{Additional Analytical Solutions}
\label{app:additional_analytics}
Following existing analytical spectral models from literature\cite{Pope00} with $D=1$ and power law $1 < \beta < 3$,
\begin{align}
    \label{eqn:model_spec_G14_2}
    \mathcal{E}_1(k) = \frac{2}{\pi} \bracket{\frac{A^2}{A^2 + k^2}}^{\beta/2} ,
\end{align}
the equivalent spectrum is,
\begin{align}
    \label{eqn:model_spec_bessel}
    \begin{split}
    \mathcal{B} \widetilde{\mathcal{E}}^{S}_1(k_{\mathrm{e}}) = &\frac{C}{b} \left(\frac{A b}{2 k_{\mathrm{e}}}\right)^{H + 1} \bigg[\\
    &\frac{Ab}{k_{\mathrm{e}}} I_{H - 1}\left(\frac{Ab}{k_{\mathrm{e}}} \right) - \frac{Ab}{k_{\mathrm{e}}} I_{-(H - 1)}\left(\frac{Ab}{k_{\mathrm{e}}} \right)\\
    &+ \frac{Ab}{k_{\mathrm{e}}} I_{H + 1}\left(\frac{Ab}{k_{\mathrm{e}}} \right) - \frac{Ab}{k_{\mathrm{e}}} I_{- (H + 1)}\left(\frac{Ab}{k_{\mathrm{e}}} \right)\\
    &+ 2 H I_{H}\left(\frac{Ab}{k_{\mathrm{e}}} \right) - 2 H I_{- H}\left(\frac{Ab}{k_{\mathrm{e}}} \right) \bigg] ,
    \end{split}
\end{align}
where $C = \frac{\sqrt{\pi}}{\sin(\pi H) \Gamma\bracket{H + \frac{1}{2}}}$, $I_\nu(\cdot)$ is a modified Bessel function of the first kind of $\nu$-th order, and $\beta = 1 + 2H$ for $0 < H < 1$. Whilst the term $\bracket{\frac{Ab}{2 k_{\mathrm{e}}}}^{H + 1}$ does not have the correct behaviour (it is only $\sim k_{\mathrm{e}}^{- H - 1}$), the remaining $\sim k_{\mathrm{e}}^{-H}$ term can be seen by expanding the series representation of the modified Bessel functions
\begin{align}
    I_\nu\bracket{\frac{A b}{k_{\mathrm{e}}}} = \sum_{m=0}^{\infty} \frac{1}{\Gamma\bracket{m + 1} \Gamma\bracket{m + \nu + 1}} \bracket{\frac{A b}{2 k_{\mathrm{e}}}}^{2m + \nu} ,
\end{align}
up to terms of $\mathcal{O}\bracket{k_{\mathrm{e}}^{-1}}$. For $k_{\mathrm{e}} \gg 1/A$, the terms in the square brackets are dominated by $\sim k_{\mathrm{e}}^{-H}$, which primarily stems from the above series representation of the term $2HI_H\bracket{\frac{Ab}{k_{\mathrm{e}}}}$. This gives us the remaining term to match the expected power law behaviour of $\sim k_{\mathrm{e}}^{-2H - 1}$.

\section{Dimensional Considerations}
\label{app:dimensional_considerations}

\subsection{\red{$D$-dimensional Spectra}}
\label{app:d_spectra}
\blue{We define the angle integrated spectrum $\Earg$ by first defining the modal spectrum
\begin{align}
    \label{eqn:fourier_spectrum_defn}
    E_{D}(\vec{k}) = \frac{1}{\bracket{2\pi}^D} \int_{-\infty}^{\infty} \avg{s_{D}(\vec{x}) s_{D}(\vec{x} + \vec{\ell})}_{\vec{x}} e^{-i \vec{k} \cdot \vec{\ell}} \d^{D} \vec{\ell},
\end{align}
with $\avg{\cdot}_{\vec{x}}$ averaging over spatial coordinate $\vec{x}$ only. The angle-integrated spectrum is then obtained by integrating over a spherical shell $\mathcal{S}(k)$ of radius $k = \abs{\vec{k}}$,
\begin{align}
    \label{eqn:integrated_spectrum_from_modal}
    \mathcal{E}_{D}(k) = \int E_{D}(\vec{k}) \d \mathcal{S}(k),
\end{align}
where
\begin{align}
    \label{eqn:unit_sphere_factor}
    \Omega_D &= \frac{2 \pi^{D/2}}{\Gamma(D/2)},
\end{align}
is the surface of the unit-sphere in $D$-dimensions where $\Omega_1 = 2$, $\Omega_2 = 2 \pi$, and $\Omega_3 = 4 \pi$. The angle-integrated spectrum, $\Earg$ can be expressed in terms of the angle-averaged autocorrelation function,
\begin{align}
    \label{eqn:omni_from_acf_transform}
    \mathcal{E}_{D}(k) = \frac{\Omega_{D}^2}{\bracket{2\pi}^D} \int_0^\infty \overline{R}_{D}(\ell) \mathcal{T}_{D}(k\ell) \bracket{k \ell}^{D-1} \d \ell,
\end{align}
where
\begin{align}
    \label{eqn:transform}
    \mathcal{T}_D(x) &= \frac{\Gamma(D/2) \mathcal{J}_{D/2 - 1}(x)}{(x/2)^{D/2 - 1}},
\end{align}
encapsulates the angle-averaging/integration, and integral transformation process for $\overline{R}_{D}(\ell)$ and $\mathcal{E}_{D}(k)$. The functions $\Gamma(\cdot)$, and $\mathcal{J}_n(x)$ are the (complete) Gamma function and the Bessel function of the first kind of $n$-th order respectively. For common $D$ values, we get the expected transforms\citep{Bracewell86, Stutzki.etal98}: $\mathcal{T}_1(x) = \cos(x)$, $\mathcal{T}_2(x) = \mathcal{J}_0(x)$, and $\mathcal{T}_3(x) = \frac{\sin(x)}{x}$. In other words, $\mathcal{E}_{D}(k)$ is the appropriate $D$-dimensional Fourier transformation.}

\red{Note, our convention of $\mathcal{E}_{D}(k)$ is twice the usual turbulence defintion, \ie{}we have $\int_0^\infty \mathcal{E}_{D}(k) \d k = \avg{s_{D}(\vec{x})^2}$ and $\mathcal{E}_{D}(k) \d k$ would be twice the energy per unit mass. It is a simple procedure to multiply the middle equation of \autoref{eqn:sf_fundamental_theorem} by a factor of $2$ and propagate to the below equations to obtain the usual turbulence theory convention.}

\subsection{\red{Sliced and Projected Data}}
\label{sec:sliced_and_projected_data}

\red{In a lot of practical cases, observations of phenomena are restricted to only 1D slices, 2D projections, or 2D slices. These restrict the type of spectrum we obtain and could introduce additional artefacts and aliasing to the spectral estimates.}

\red{In the case of the solar wind (\autoref{sec:solar_wind_example}), measurements are performed by sensors moving in relation to the plasma rest frame. These sensors generate a time-series of data: such as magnetic field measurements $\vec{b}(t)$. We can construct the structure function and autocorrelation functions as,
\begin{align}
    S(\tau) &= \avg{\abs{\vec{b}(t) - \vec{b}(t + \tau)}^{2}},\\
    R(\tau) &= \avg{\vec{b}(t) \cdot \vec{b}(t + \tau)}
\end{align}
where $\avg{\cdot}$ averages over $t$, $\abs{\cdot}$ is the vector norm, and we assume stationarity. We can also Fourier analyze this data by applying the Fourier transform to $R_{1}(\tau)$, giving a PSD $\mathcal{P}(\omega)$ that depends on frequency. Following the steps in \autoref{sec:method}, we see that,
\begin{align}
    \int_0^\infty \frac{\d S(\tau)}{\d \tau} \d \tau &= \int_0^\infty \mathcal{P}(\omega) \d \omega,
\end{align}
leads to the following approximation\cite{Thepthong.etal23}
\begin{align}
    \label{eqn:equivalent_spectrum_time_dependent}
    \frac{1}{2} \frac{1}{b} \tau^2 \frac{\d S(\tau)}{\d \tau} \approx \mathcal{B} \mathcal{P}(\omega).
\end{align}}

\red{By assuming that the bulk flow is sufficiently fast, the temporal variations of $\vec{b}(t)$ at a fixed location ($\approx 1\,\mathrm{au}$) can be interpreted as spatial variations by $\vec{x} = - \vec{U} t$ (Taylor's frozen-in hypothesis) \cite{Frisch95, Verma22} where $U \approx 400\,\mathrm{km/s}$ is the average bulk flow speed of the slow solar wind \cite{McComas.etal00}. Using Taylor's frozen-flow hypothesis, we can interpret the time/frequency functions, $S(\tau)$, $R(\tau)$, $\mathcal{P}(\omega)$, as (approximations to) spatial functions with the lag $\ell_1$ in the direction of the bulk flow $U$,
\begin{align}
    S_{3}(\ell_1 = - U \tau, 0, 0) &\approx S(\tau),\\
    R_{3}(\ell_1 = - U \tau, 0, 0) &\approx R(\tau),
\end{align}
and the Fourier transform of $R_{3}(\ell_1, 0, 0)$ corresponds to the ``reduced'' spectrum\cite{Oughton.etal15},
\begin{align}
    \label{eqn:reduced_spectrum}
    E_{\mathrm{red}}(k_1) = \int_{-\infty}^{\infty} \int_{-\infty}^{\infty} E(k_1, k_2, k_3) \d k_2 \d k_3.
\end{align}}

\red{The exact association to the full spectrum requires careful consideration and futher assumptions \eg{}whether the turbulence is restricted to 2D planes, or is fully 3D \cite{Fredricks.Coroniti76}. However, assuming fully 3D isotropic turbulence, we obtain,
\begin{align}
    \widetilde{\mathcal{E}}_{\mathrm{red}}^{S}(k_{\mathrm{e,1}}) \equiv \frac{1}{2} \frac{1}{b} \ell_1^2 \frac{\partial S_{3}(\ell_1, 0, 0)}{\partial \ell_1} \approx \mathcal{B} E_{\mathrm{red}}(k_1)
\end{align}
where $k_{e,1} = b/\ell_1$ is the equivalent wavenumber associated with $k_1$.}

\red{From \autoref{eqn:reduced_spectrum}, it is easy to see that there is a form of aliasing where $E_{\mathrm{red}}(k_1)$ is the energy associated with all wavevectors that have the same $k_1$ component. For example, some model functions from \autoref{tab:b_factors} for $E_{\mathrm{red}}(k_1)$ are shown in \autoref{tab:b_factors2}. These functions are largest at $k_1=0$. We also show the associated $\widetilde{\mathcal{E}}_{\mathrm{red}}^{S}(k_{\mathrm{e,1}})$ in \autoref{tab:b_factors2} and can see that $\widetilde{\mathcal{E}}_{\mathrm{red}}^{S}(k_{\mathrm{e,1}})$ has a peak not at the origin -- $\widetilde{\mathcal{E}}_{\mathrm{red}}^{S}(k_{\mathrm{e,1}})$ does not appear to exhibit such aliasing as with $E_{\mathrm{red}}(k_1)$ \cite{Davidson15, Chola.Chakraborty25}. Nevertheless, as with \autoref{eqn:equivalent_spectrum}, we can investigate cases in which the approximation holds true for appropriate biases $b$ and $\mathcal{B}$ for a range of $\ke$.}

\begin{table}
    \centering
    \begin{tabular}{ c c c }
        \hline
        \hline
        $\dfrac{\overline{R}_{3}(\ell)}{A}$ & $\dfrac{E_{\mathrm{red}}(k_1)}{A}$ & $\dfrac{ \widetilde{\mathcal{E}}_{\mathrm{red}}^{S}(k_{\mathrm{e,1}})}{A}$ \\ 
        \hline
        \hline
        $\dfrac{\bracket{3L - \ell} e^{-\ell/L}}{L}$ & $\dfrac{2 L}{\pi} \bracket{L^2 k_1^2 + 1}^{-1}$ & $\dfrac{2b^2 e^{- \frac{b}{L k_{\mathrm{e,1}}}}}{L k_{\mathrm{e,1}}^2}$ \\
        $\dfrac{\bracket{3L^2 - \ell^2} e^{-\ell^2/L^2}}{L^2}$ & $\dfrac{L}{\sqrt{\pi}} e^{-L^2 k_1^2 / 4}$ & $2b^2 e^{- \frac{b^2}{L^2 k_{\mathrm{e,1}}^2}}$\\
    \end{tabular}
    \caption{\textbf{Analytical solutions for reduced spectra using different model autocorrelation functions}. \red{Shown are: the reduced spectra $E_{\mathrm{red}}(k_1)$ and the equivalent structure function estimator $\widetilde{\mathcal{E}}_{\mathrm{red}}^{S}(k_{\mathrm{e,1}})$ using these different forms of the autocorrelation function $\overline{R}_{3}(\ell)$.}}
    \label{tab:b_factors2}
\end{table}

\red{Above, we see that a slice of data produces projections of the Fourier spectrum. Now, we look at projections of data which are common in astronomical observations of turbulence such as the interstellar medium (\autoref{sec:ISM_example}). For a 3D scalar field $s_{3}(x,y,z)$ the plane-of-sky (POS) observation $p(x,y)$ is the projection along the line-of-sight (LOS) $z$-axis,
\begin{align}
    p(x,y) = \int_{\mathrm{LOS}} s_{3}(x,y,z) \d z,
\end{align}
where we have assumed there are no additional emissivity or modulation factors inside the integrand. From the \emph{projection-slice theorem}, the spectrum of $p(x,y)$ -- the plane-of-sky spectrum $E_{\mathrm{POS}}(k_x, k_y)$ -- is proportional to a slice of the full 3D spectrum of $s_{3}(x,y,z)$ at LOS wavenumber $k_z = 0$,
\begin{align}
    E_{\mathrm{POS}}(\vec{k}_{2}) \propto E_{3}(\vec{k}_{2}, k_z = 0),
\end{align}
where $\vec{k}_{2} = (k_x, k_y)$. By integrating wavenumber shells $\kappa_2 = \abs{\vec{k}_{2}}$ in the observed 2D Fourier plane (following \autoref{eqn:integrated_spectrum_from_modal}) we obtain $\mathcal{E}_{\mathrm{POS}}(\kappa_{2}) \ne \mathcal{E}_{3}(k)$ where $\mathcal{E}_{3}(k)$ is the angle-integrated spectrum of $s_{3}(x,y,z)$.}

\red{We assume that $\ESarg$ approximates the angle-integrated Fourier spectrum that is computed on the \emph{observed} data \ie{}$\mathcal{E}_{\mathrm{POS}}(\kappa_2)$, $\mathcal{P}(\omega)$. Conversion from the observed spectrum to the spectrum representing the dynamics of the system, \eg{}$\mathcal{E}_{3}(k)$, is a separate problem altogether.}

\section{\red{Relation to Signature Functions}}
\label{app:signature_function_relations}

\red{We use the integrands of \autoref{eqn:esf_integrals} to motivate \autoref{eqn:equivalent_spectrum}. This, of course does not mean we assume point-wise equality of the integrands of \autoref{eqn:esf_integrals}. If we wished to assume point-wise (approximate) equality of the integrands, an equivalent expression would be:
\begin{align}
    \label{eqn:delta_function_integrands}
    \frac{1}{2} \frac{1}{b} \ell^2 \int_0^\infty \delta(\ell - r) \frac{\d \overline{S}_{D}(r)}{\d r} \d r &\approx - \int_0^\infty \delta(k - \ke) \mathcal{E}_{D}(k) \d k.
\end{align}
From $\frac{\d \Theta(x)}{\d x} = \delta(x)$, where $\Theta(x)$ is the Heaviside step-function, \autoref{eqn:delta_function_integrands} is
\begin{align}
    \label{eqn:signature_function_derivative_filter}
    \frac{1}{2} \frac{1}{b} \ell^2 \frac{\partial}{\partial \ell} \int_0^\ell \frac{\d \overline{S}_{D}(r)}{\d r} \d r &\approx - \frac{\partial}{\partial \ke}\int_{\ke}^{\infty} \mathcal{E}_{D}(k) \d k.
\end{align}
Intrinsic to the process in \autoref{eqn:signature_function_derivative_filter} is the filtering:
\begin{align}
    \label{eqn:SF_PSD_filter_approx_assumption}
    \int_0^\ell \frac{\d \overline{S}_{D}(r)}{\d r} \d r \approx \int_{\ke}^{\infty} \mathcal{E}_{D}(k) \d k,
\end{align}
\ie{}the assumption that the structure function is well approximated by the sum of energy of all wavenumbers $k \ge \ke$,
\begin{align}
    \overline{S}_{D}(\ell) \approx \int_{\ke}^{\infty} \mathcal{E}_{D}(k) \d k.
\end{align}
In this sense, $\ke$ could be referred to as a ``cut-off wavenumber'' rather than an ``equivalent wavenumber''.
}


\red{It is worth noting from \autoref{eqn:spectral_representation}, that really, there are higher-order contributions to $\overline{S}_{D}(\ell)$\cite{Davidson15}, \eg{}ignoring order-unity constants and Taylor expanding $1 - \mathcal{T}_{D}(k\ell)$ for small $k$,
\begin{align}
    \overline{S}_{D}(\ell) \approx \int_{\ke}^{\infty} \mathcal{E}_{D}(k) \d k + \ell^2 \int_0^{\ke} k^2 \mathcal{E}_{D}(k) \d k + \dots.
\end{align}
In other words, $\overline{S}_{D}(\ell)$ is the sum of energy in scales $\le \ell$ and the enstrophy in scales $\ge \ell$ (plus higher-order corrections). In \autoref{sec:filter_function}, we see that $\ESarg$ is \emph{not} described by filters with sharp cutoffs.}

\red{A similar branch of literature worth mentioning is the \emph{signature function}\cite{Townsend76, Davidson.Pearson05, Davidson15, Hamba15, Hamba18, Chola.Chakraborty25}, $\mathcal{V}(\ell)$, which is intended to describe the turbulent energy distribution in real, rather than $k$-space with the following requirements:
\begin{description}
    \item[Requirement 1\label{point1}] $\mathcal{V}(\ell) \ge 0$,
    \item[Requirement 2\label{point2}] $\int_0^\infty \mathcal{V}(\ell) = \avg{s_{D}(\vec{x})^2}$,
    \item[Requirement 3\label{point3}] for a random distribution of eddies of size $\sim \ell_e$, the corresponding $V(\ell)$ has a clear peak around $\ell \sim \ell_e$.
\end{description}
No function has been found to meet all of these requirements, so usually, \nameref{point1} is relaxed to $\int_0^\ell \mathcal{V}(r) \d r \ge 0$ \cite{Davidson15, Chola.Chakraborty25}.}

\red{It has been noted that the relationship of the signature function to the angle-integrated spectrum is,
\begin{align}
    \label{eqn:sig_E_relation}
    \ell \mathcal{V}(\ell) \approx \sqbracket{k \Earg}_{\ke}.
\end{align}
which is similar to \autoref{eqn:equivalent_spectrum_relationship} upon re-arranging, for $\mathcal{V}(\ell) = \frac{\d \overline{S}_{D}(\ell)}{\d \ell}$, $\ke = b/\ell$, \red{and assuming $\mathcal{B}=1$ (point-wise equality)}. The estimate, \autoref{eqn:sig_E_relation}, is poor when $\mathcal{E}_{D}(k)$ exhibits steep gradients -- which often occur outside the turbulence inertial range -- and facilitates in the proposed constraint of the valid region of $\eta \ll \ell \ll L_I$ where $\eta$ is the Kolmogorov microscale and $L_I$ is the integral scale\cite{Davidson15}. This remark is backed by examination of $\mathcal{V}(\ell)$ and how it represents models with energy-containing ranges (like \autoref{sec:analytical_solutions_of_model_correlation_functions} and \autoref{sec:exp_analytical_bias} for $D=3$) \cite{Davidson15}.}

\red{Our approach (see, \autoref{eqn:equivalent_spectrum_relationship}) does \emph{not} assume point-wise equality: $\mathcal{B}$ accounts for this.} \red{Practically, $\mathcal{B}$ allows for corrections based on local gradients and (in the context of \autoref{eqn:sig_E_relation}) reduces the error in the approximation. This correction factor can be modelled in several ways: the first requiring additional \emph{a priori} assumptions about the shape of $\mathcal{E}_{D}(k)$ (a parametric-style approach, \eg{}\autoref{eqn:meijerG_bias}), and another that assumes local power-law behaviour (a non-parametric style approach, \autoref{sec:powerlaw_approx}). Of course, $\mathcal{B}$ is not universally applicable, and is merely sought to reduce the error for some range of $\ell$.}

\section*{Author Contributions}
M.A.B. contributed formal analysis, investigation, methodology, software, visualization, writing -- original draft

S.O. contributed methodology, supervision, writing - review \& editing

T.N.P. contributed conceptualization, methodology, supervision, writing - review \& editing

Y.C.P. contributed funding acquisition, methodology, supervision, writing - review \& editing


\begin{acknowledgments}
Research supported by the Marsden Fund Council from  NZ Government funding, managed by Royal Society Te Apārangi (E4200).

We thank the following for making their datasets accessible:
\emph{Wind} MFI and 3DP instrument teams and NASA GSFC's Space Physics Data Facility\footnote{\url{https://spdf.gsfc.nasa.gov/}}, Clark et al.\cite{Clark.etal21, Clark.etal23} for the LMC dataset, Cardesa et al. \cite{Cardesa.etal17} for the incompressible hydrodynamic simulation.

The authors wish to acknowledge the use of Victoria University of Wellington's High Performance Compute Cluster Rāpoi as part of this research.

The authors wish to acknowledge the use of New Zealand eScience Infrastructure (NeSI) high performance computing facilities as part of this research. New Zealand's national facilities are provided by NeSI and funded jointly by NeSI's collaborator institutions and through the Ministry of Business, Innovation \& Employment's Research Infrastructure programme.\footnote{\url{https://www.nesi.org.nz}}

We thank D. Wrench for providing additional manuscript review.

We thank the anonymous referees for helpful suggestions that have improved the presentation and quality of this paper.
\end{acknowledgments}

\bibliography{References}{}

\end{document}